\documentclass[sn-mathphys-num]{sn-jnl}

\usepackage{graphicx}%
\usepackage{multirow}%
\usepackage{amsmath,amssymb,amsfonts}%
\usepackage{amsthm}%
\usepackage{mathrsfs}%
\usepackage[title]{appendix}%
\usepackage{xcolor}%
\usepackage{textcomp}%
\usepackage{manyfoot}%
\usepackage{booktabs}%
\usepackage{listings}%
\usepackage{lineno}%

\theoremstyle{thmstyleone}%
\theoremstyle{thmstyletwo}%

\theoremstyle{thmstylethree}%

\begin{document}

\title[World Cities' Land Footprint]{Removing Population Size, World Cities Leave on Land a Footprint of Wealth}


\author[1]{\fnm{Victor} \sur{Vignolles}}\email{victorvignolles@gmail.com}

\author*[1]{\fnm{R\'{e}mi} \sur{Lemoy}}\email{remi.lemoy@univ-rouen.fr}


\affil*[1]{\orgdiv{Geography Department, University of Rouen}, \orgname{IDEES Laboratory, UMR 6266 CNRS}, \orgaddress{\street{Rue Lavoisier}, \city{Mont-Saint-Aignan}, \postcode{76821}, \country{France}}}








\abstract{Urban expansion builds on precious arable land \cite{seto2011,bren2017}, affecting ecosystems and health \cite{mcdonald2020,simkin2022,seto2012,schwela2000,WHO2020}, despite worldwide measures to alleviate its impacts, in small and large cities. Nonetheless, the link between urban extent and population size is still unclear \cite{batty2011,ribeiro2023}. Here we uncover a scaling law governing built-up land in 1800+ global urban areas. We observe that world cities have homothetic (or isometric) radial land use profiles, and that built-up footprint is proportional to total population. This spatial scaling law is important for the understanding and the definition of urban areas, to help build a science of cities and make them more sustainable. It implies that small and large cities have similar internal structures, and that built-up area per capita is constant across city sizes. The remaining variations are very strongly correlated to wealth measured by gross domestic product per capita, suggesting a pick between economic development and sustainabiliy.}




\maketitle

Human societies are increasingly urban at the global scale. Urbanization is associated with economic, scientific, and societal progress \cite{glaeser2012} for humanity, and also important consequences on the environment, such as climate change or losses of biodiversity and arable land, which are linked to urban land use \cite{seto2011,bren2017,mcdonald2020,simkin2022,seto2012,schwela2000,WHO2020}. Cities participate in the biodiversity crisis in particular through animal habitat fragmentation. Indeed, urban forms follow fractal patterns \cite{white1993}, which means that they affect vast areas, although urban land represents only around 1.47\% of global land area in 2015 (1.05\% in 1990) \cite{CIESIN2013}. Climate change is also linked to cities as they concentrate human activities and associated pollutions, for instance the emissions of greenhouse gases. These affect deeply the global climate, while the urban heat island effect, linked to urban land use, changes the local intra-urban climate \cite{zhou2017}. Actually, soil sealing is an issue probably first of all because land which is taken by urbanization is usually rich and easily accessible arable land \cite{bren2017}. Its loss affects agricultural opportunities and can be linked with deforestation and zoonoses. Arable land is one of many finite, non-renewable (in reasonable time scales) planetary resources which need to be used sparingly \cite{rome,iea2023}. Indeed, once a piece of land has been urbanized, covered e.g. by a building or road, the soil dies quickly. If it is then unsealed, it can become fertile again only after a very long time, through natural processes. Soil sealing is also a major driver of flood risk \cite{wheater2009}.
In this context, the spatial aspects of urban land use elude a precise quantitative description, despite a huge body of literature on the statics and dynamics of cities. City definition itself is still a rather open problem, especially at the global scale. Hence, there is a need for robust empirical stylized facts to guide the emerging urban science. Here we study the distribution of urban land use within global cities, and its evolution with city size. We show that it can guide us for defining cities, and help us determine whether smaller or larger cities are more parsimonious in their use of land.

We note first that the generic shape of cities is circular. They are usually organized around a city center, where human activities are especially concentrated. This is associated with higher buildings, high levels of built up and artificial land, of residential population density, employment and urban amenities. It is the reason why quantitative spatial models of urban structure started with a radial approach to cities \cite{vonthunen,alonso}. Here, in order to find a generic urban structure, we focus only on the most essential spatial characteristics of cities, which might be observed for very different entities worldwide. We observe that the urban radial structure has such a generic character.

Second, we know that cities exist in a wide variety of scales. The distribution of their size has been much studied and linked to Zipf's law \cite{nitsch2005}. Many attributes of cities, such as average income or air pollution also evolve with size, and their study constitutes the domain of urban scaling laws \cite{pumain2004,batty2008,bettencourt2013}. This study actually started in biology, where scaling laws for organisms link their mass to other attributes, such as heart rate or bone section \cite{brown2000}. Scaling laws in biology are grounded on the obvious definition of mass, which is a good measure of organisms' size. For urban scaling laws, total population $N$ is usually considered an equivalent, although it is dimensionless. It is the size parameter of cities which is best defined, because most of the urban population is concentrated near the city center, and the total population is not too sensitive to city definition. One of the important differences between biological organisms and cities is indeed that humans who animate cities are largely independent, which means that the city does not have as much coherence, as a whole, as a biological organism. This explains why the question of city definition is still open after decades of research on the subject.

Hence we study here how center-periphery profiles of urban land use evolve with city size measured by population N. This links intra-urban study (radial analysis) to inter-urban study (scaling laws) and can be seen as a study of ``cities as systems within systems of cities" \cite{berry1964}. More precisely, we study the radial profiles of urban built-up land in the 1800 largest urban areas of the world, with more than 300,000 inhabitants each. And we find that these center-periphery profiles present striking regularities, within cities and across city sizes and countries of the world. This provides a new robust stylized fact, much needed for the advancement of urban science. 

\section*{Uniformity of land use in world cities}

\begin{figure}[h]
\centering
\includegraphics[width=0.98\textwidth]{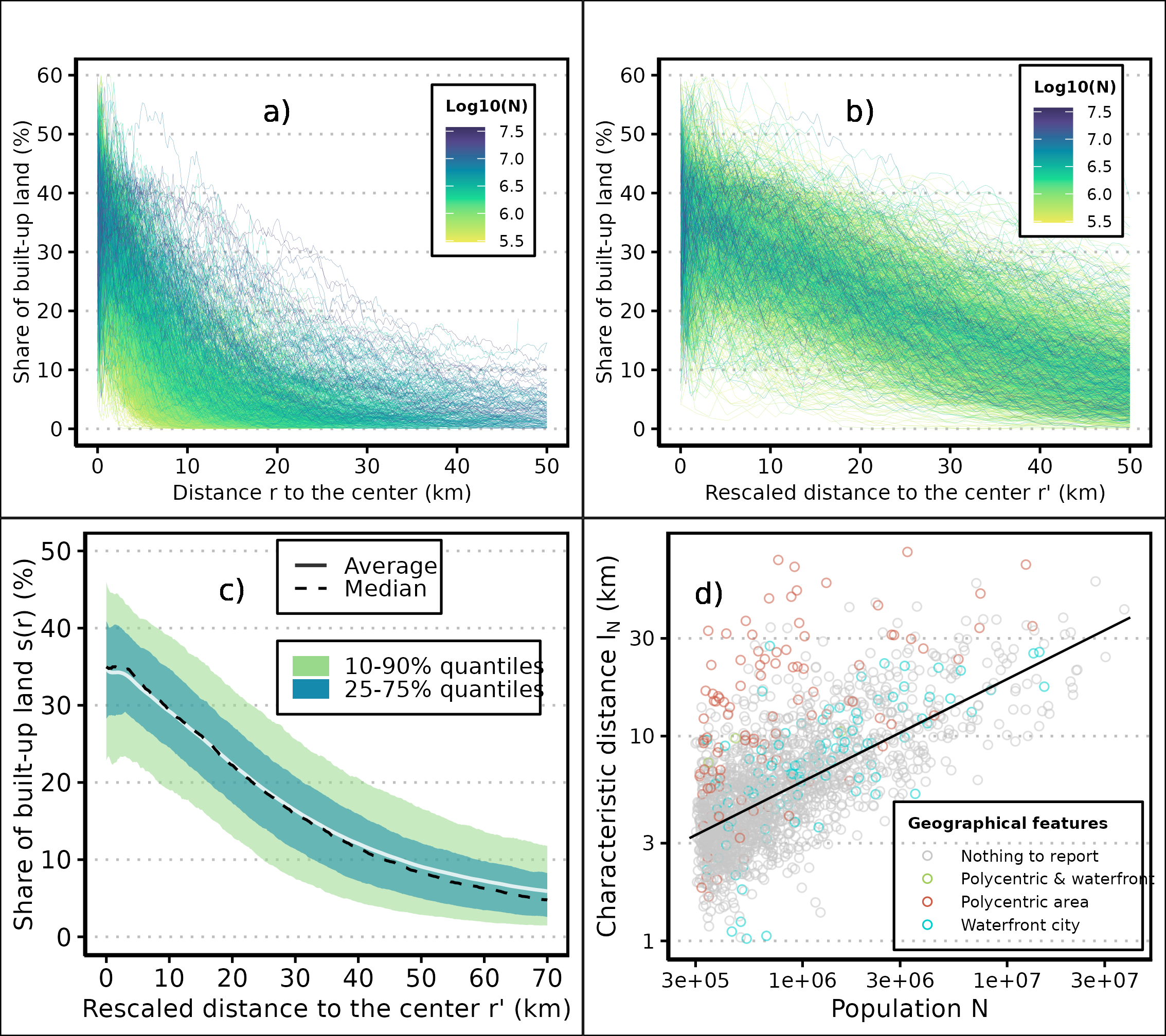}
\caption{Urban radial structure of built-up land in 2020. Share of built-up land for the 1800+ largest global cities as a function of the a) non-rescaled distance $r$ and b) rescaled distance $r'$ to the center. c) Statistics on the rescaled profiles, showing a clear common trend. d) Characteristic radius $l_N$ as a function of population $N$, with a line $l_N=l_1 \sqrt{N}$, where $l_1=6$ m, used as a guide to the eye (see Supp. Table 1 and Supp. Fig. 11).}
\label{fig1_resc}
\end{figure}

Using radial (center-periphery) profiles allows us to project (or average) the two-dimensional urban land on only one dimension, the distance $r$ to the city center (Fig. \ref{fig1_resc}a)), which is convenient to assess how the share of built-up land $s(r)$ decreases when the distance to the center $r$ increases. Then we use two methods to remove the effect of size and identify the scaling law governing these urban land use profiles. The first one, illustrated on Fig.\ref{fig1_resc}b)-c), consists in rescaling the distance to the center proportionally to the square root of city population $\sqrt{N}$. The rescaled distance $r'$ is given by $r'=r.k_N$, with $k_N=\sqrt{N_{\text{Tokyo}}/N}$ the rescaling factor, where we use Tokyo, the largest city in the world with population $N_{\text{Tokyo}} \simeq 3.10^7$ inhabitants, as a reference (without loss of generality). For a city 4 times smaller than Tokyo such as London, ($N_{\text{London}}=N_{\text{Tokyo}}/4$), the distance is multiplied by a rescaling parameter $k_{\text{London}}=\sqrt{4}=2$. In other words, the area of each city is stretched proportionally to its population $N$. This provides a data collapse of the built-up land profiles of global cities, and a generic radial profile appears (Fig. \ref{fig1_resc}b)-c)). This profile has an exponential form and a characterestic scale (it is not scale-free). Our second method consists then in an exponential fit of the radial built-up land profile $s_N(r)$ of each city, following $s_N(r)=a_N \exp (-r/l_N)$. We find that the share of built-up land in the city center $a_N$ is rather constant, around 35\% (see Supp. Info.). And the characteristic distance (scale) $l_N$ of these exponentially decreasing profiles scales like the square root of population $l_N \sim l_1.\sqrt{N}$ (Fig. 1(d)), where $l_1\simeq 6$m is the characteristic radius of a theoretical city with $N=1$ inhabitant (see Supp. Info.). The results of both methods are clearly consistent and robust. Built-up land is strikingly similar in world cities, once the effect of size is taken off. And this effect is simple and geometrical. As the characteristic radius $l_N$ of a city is proportional to the square root of its population $l_N\sim \sqrt{N}$, the built-up area $S$ of a city is proportional to its population  $S\sim l_N^2\sim N$. This means that the built-up area per capita is constant, and that inhabitants of small or large cities use just as much built-up land on average. It also means that efforts to curb urban sprawl and preserve arable and natural land are needed equally in small and large cities.
We note that the relation between area $S$ and population $N$ of cities has already been studied rather extensively, but without a conclusive result. The exponents coined in the literature \cite{batty2008,batty2011,ribeiro2023} range between 2/3 and 1, the value we find here. It is a fundamental relation, which is the starting point of different modelling approaches \cite{bettencourt2013,ribeiro2023}. Here we reach a definite conclusion, at the global scale, for two reasons. First, we focus on land use only, which is probably the most regular morphological aspect of cities. Second, our radial scaling methods allow us to exploit the generically circular structure of cities and extract the size effect efficiently.
Furthermore, we obtain a simple universal formula governing built-up land in world cities $s_N(r)=a \exp (-r/(l_1 \sqrt{N}))$, with $a \simeq 35\%$ and $l_1\simeq 6$m. This is a precious stylized fact which can be a guide for city definition regarding environmental impacts such as losses of biodiversity, land take or urban heat island, which are all linked to the urbanized area of cities \cite{seto2011,simkin2022,seto2012,zhou2017}. We note that very similar results can likely be obtained for artificial or impervious urban land \cite{tobler1969,guerois2008,lemoy2020,lemoy2021}, even though datasets at the global scale are not as reliable yet.

\begin{figure}[h]
\centering
\includegraphics[width=0.98\textwidth]{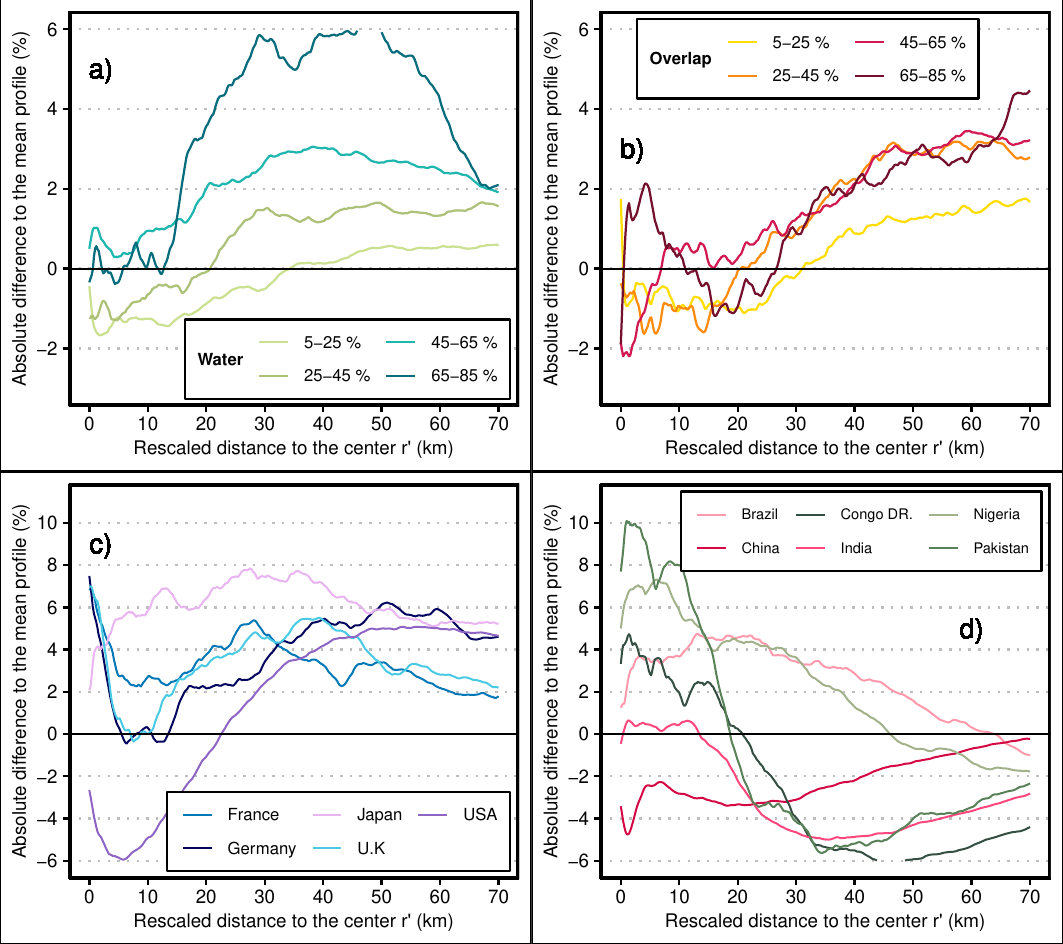}
\caption{Deviation from the mean rescaled built up land profile, depending on geographical and economic factors. a) Waterfront cities, b) Polycentric areas, c) 5 countries with high GDP per capita, d) 3 of the BRICS countries (Brazil, China, India) and 3 developing countries (Congo DR., Nigeria, Pakistan).}\label{fig2_dev}
\end{figure}

We note also that water bodies and polycentric urban regions have some influence on the radial profiles of built-up land studied here. We first need to state that water bodies have been removed from our analyses. Since land under water is not built-up, the share of built-up land is computed over emerged land only, which is a rather effective way to treat coastal cities. Indeed, to put it simply, cities which include a large area of water bodies, such as coastal cities, present radial profiles which are quite similar to more continental cities, once water bodies are removed from the analysis. This can be seen on Figure \ref{fig1_resc}d), where waterfront cities follow the general trend, and on Figure \ref{fig2_dev}a), where deviations due to water bodies are rather small (a few percentage points, even for cities having most of their area covered by water). We note however that there is a trend in these deviations as  cities with more water bodies have more built-up land in their periphery, on average. On the other hand, polycentric urban regions introduce more fluctuations on Figure \ref{fig1_resc}d), and an understandably higher characteristic distance on average, although the deviations they introduce in radial profiles on Figure \ref{fig2_dev}b) are also rather small.

\section*{Wealth, hidden behind the uniformity of land use patterns}

Our scaling analyses also suggest to define a size-independent indicator of urban extent per capita, which we call Urban Built Land Index (UBLI), as $\text{UBLI}=a_N.l_N^2/N$. Since the share of built-up land in the city center $a_N$ is roughly constant (see Supp. Info.), this index is closely related to the residual of the regression on Fig. \ref{fig1_resc}d) (see Supp. Fig. 14a)). We note also that this UBLI has a value which is on average $l_1^2 \simeq 34$ m$^2$ if we forget $a_N$ (roughly constant). This distance $l_1 \simeq 6$ m (the characteristic length of a theoretical city with $N=1$ inhabitant) is a new fundamental constant of cities, and $l_1^{-2} \simeq 29,000$ inhabitants/km$^2$ gives a characteristic scale of population density. Contrary to studies of scaling laws in physics or biology, this UBLI indicator is not dimensionless. This is linked to the fact that total population $N$, the size parameter, is a dimensionless measure of human life in a city (an "urban mass"), which actually measures many different phenomena studied in urban scaling analysis, related to economic and social activity, living conditions or pollution for instance. There is no equivalent in physics or biology, which also relates to our discussion above on the correspondence between city population and mass of organisms.

\begin{figure}[h]
\centering
\includegraphics[width=0.98\textwidth]{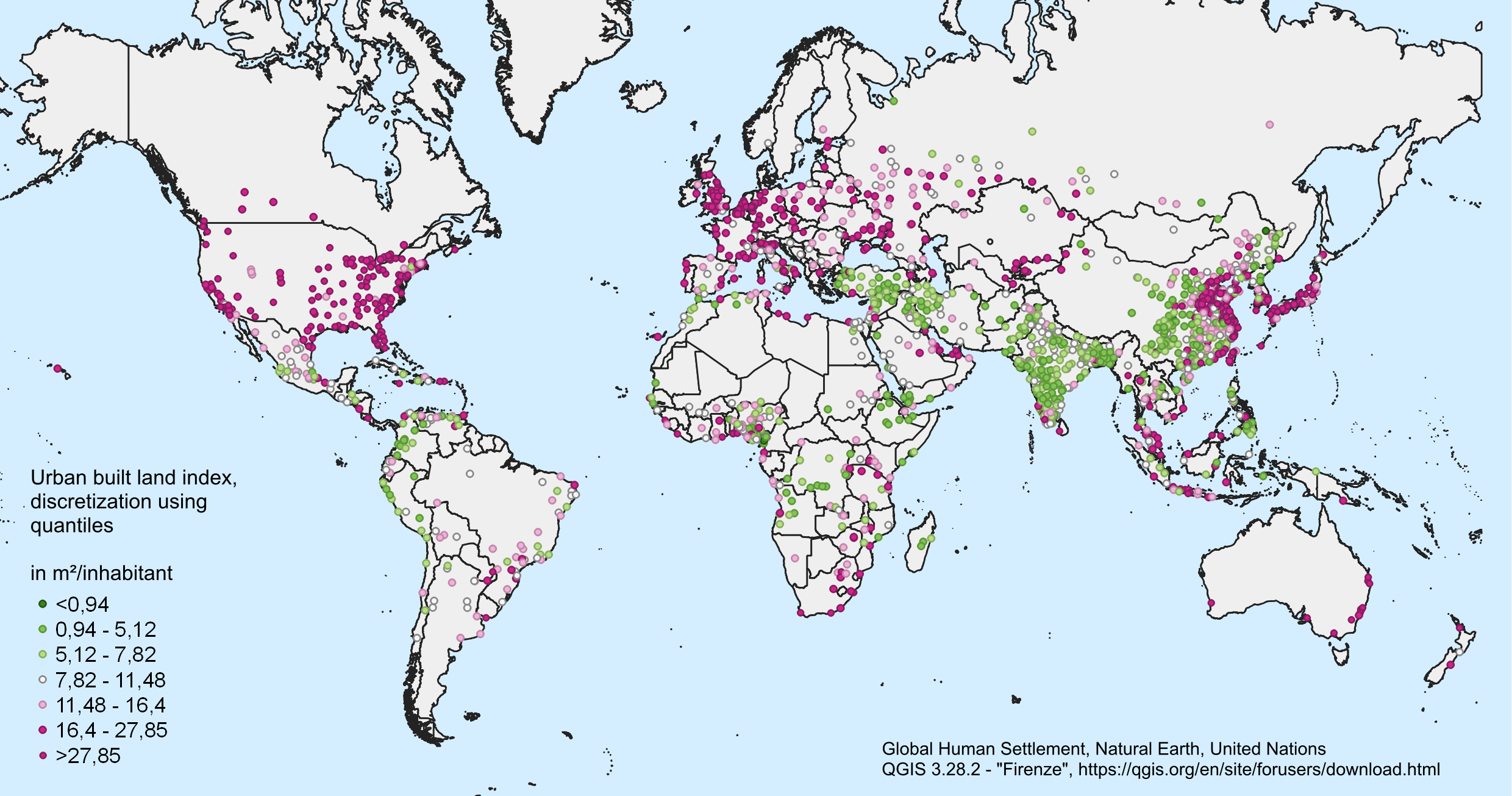}
\caption{Global map of the Urban Built Land Index (UBLI), which indicates cities' extent irrespective of their size. See Supp. Info. for zooms and a comparison with wealth.}\label{fig3_map}
\end{figure}

\begin{figure}[h]
\centering
\includegraphics[width=0.98\textwidth]{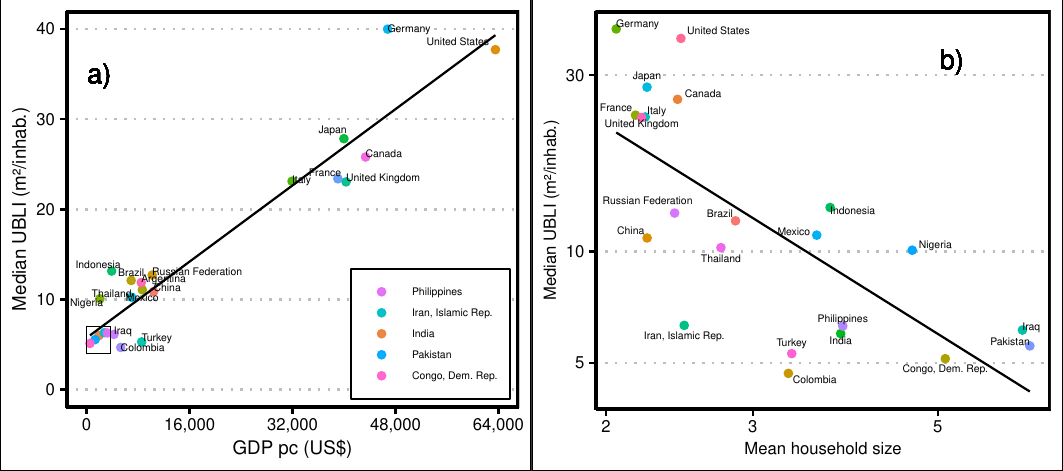}
\caption{Relationship between the median national urban built land index UBLI and a) gross domestic product per capita and b) mean household size (log-log graph).}\label{fig4_corr}
\end{figure}

This Urban Built Land Index allows us to characterize per capita urban land use independently of the size of cities. What is left of global per capita urban land variations, then, when the size effect is taken off? The answer is wealth, measured by the gross domestic product per capita (GDP pc), as shown on Figures \ref{fig3_map} and \ref{fig4_corr}. When mapping the UBLI (Fig. \ref{fig3_map}), we observe a map very reminiscent of the global map of the GDP pc (see Supp. Fig. 14). And when we relate the two variables together (GDP pc and median UBLI) for the countries with the largest number of cities, we observe indeed a surprisingly high linear correlation R=0.95 (Fig. \ref{fig4_corr}a). This level of correlation (see Supp. Fig. 13) means that the relationship between wealth and urban land use is extremely direct. Actually, it is probably two-way, since more wealth allows people to consume larger building spaces per capita, and the construction, maintenance, heating and cooling of these larger buildings generates economic activity and wealth which contributes to higher GDP pc.

We also note that wealth is a multi-faceted variable, correlated to many others such as household size (negatively, see Fig. \ref{fig4_corr} for the link between household size and UBLI) or car ownership (positively). These variables can in turn influence urban land per capita.

Turning back to rescaled radial profiles of urban land use, we can observe the influence of those variables on the intra-urban structure of cities. We see on Figure \ref{fig2_dev}c) and d) that economic development and wealth are indeed expressed in urban radial profiles, once the size effect is removed. Cities in wealthier countries have more built-up land, in particular in their periphery, where most urban land is located.

Our findings suggest that on the one hand, efforts to mitigate urbanisation and soil sealing by buildings are needed equally for small and large cities, anywhere in the world. On the other hand, efforts should be all the more important as countries and people are wealthier. As for other variables and other scales (such as carbon footprint \cite{hertwich2009,moran2018}), we find indeed an extremely direct link here between wealth and environmental impacts. In order to preserve land on earth from urbanisation and meet sustainability challenges, a general conclusion is that the goal of human economic activity and policy needs to change. The objective of increasing wealth per se (measured by GDP) is wrong, because it increases climate change, environmental pollution and the toll on planetary ressources and biodiversity. Actually, we see here that development, taken in the sense of economic growth, goes exactly against sustainability challenges. On the contrary, the aim of economic activity should be less individualistic, competitive and materialistic, and more collective, collaborative and immaterial (the enjoyment of life \cite{georgescu1979}). In this endeavor, education and research could clearly play an important role. However, this is not really the orientation which human societies are choosing now.

\section*{Methods}

\subsection*{Data}

This work relies on remotely sensed data for observing land use, which provides many opportunities for urban and environmental studies, for instance regarding urban forms \cite{zhu2019}. We use mainly the GHSL Built-up surface dataset (GHSL BUILT-S R2022A), developed by the Global Human Settlement Layer (GHSL) project of the European Commission (which is illustrated on Supp. Figs 5 and 6). This raster spatial dataset provides the global distribution of built-up areas, i.e areas covered by buildings, using a spatial generalization \cite{GHSL2020}. We choose here a dataset with a resolution of 100 m, which is sufficiently precise considering the 200m width of rings used in this study. 
The second main component of our data is the total population of cities, which we use as a scale factor. We select the database of Cities with 300,000 inhabitants or more 
from the World Urbanization Prospects of the United Nations. This represents 1860 cities and above 2.5 billion people, more than 32\% of the world population. The population for each year presents collected data (national censuses, sample surveys, civil registers), and demographic projections to fill data gaps \cite{UNWUP2018}.
Water bodies need to be considered since they can have a significant impact on urban land use distribution, and urban areas are often adjacent to rivers and coastlines \cite{grimm2008}. Consequently, we mask the GHSL built-up surface raster values by the ESRI - Garmin World Water Bodies, which contain the major oceans, seas, rivers, lakes and salt flats at a global scale. Through this process, all the pixels that intersect with any body of water are not considered in our analysis.
We also use economic and demographic data at the scale of world countries, namely GDP per capita data in 2020 provided by the World Bank, and mean househod size provided by the United Nations. The full reference of these datasets is given in the Supplemental Information.

\subsection*{Statistical and spatial analysis}

The determination of the city centers' location is an important point for our radial analysis. We focus here on city halls, due to their consistently central location. In cases where the city hall is not referenced on any reliable website, we use the location of a central structuring building, such as a central post office, a major central marketplace or bazaar, a mayoral house, a central square or a central religious monument, with the precious help of the Global Urban Centres database \cite{kilgarriff_2021}. To guarantee the robustness of this key data, all 1860 centers are returned one by one using this method. Then we perform a radial analysis for each city in our sample. We first define a maximal urban extent as a disc around each city center. This disc is of radius $D=150$ km for the urban area of Tokyo (the largest city in the sample, whose population is denoted $N_\text{Tokyo}$). The radius is smaller for other cities, according to the scaling law we uncover in our analysis. For a city of population $N$, this maximal extent is $r_{\text{max}}=D/k_N$, where $k_N=\sqrt{N_{\text{Tokyo}}/N}$, with $r$ the Euclidean distance to the center. For London, whose population is roughly 4 times smaller than Tokyo's, $k=2$ and the maximal extent is 150/2=75km (see Supp. Fig. 6). Within each city's disc, we then extract concentric rings of 200m width, in which the share of built-up land is computed. Thus we obtain our object of study, the share of built-up land $s(r)$ at any distance $r$ from the city center.
In order to make all cities comparable to Tokyo (chosen as a reference, without loss of generality), we rescale on Fig. \ref{fig1_resc}b)-c) the distance $r$ of all cities' radial profiles $s(r)$ using the rescaling factors $k_N$, thus defining a rescaled distance $r'$ to the center . We check that the exponent 1/2=0.5, corresponding to the square root of population, provides an optimal rescaling and data collapse in terms of signal-to-noise ratio (Supp. Fig. 7). Since rescaling factors $k_N$ have different real values, we use linear interpolation between the closest data points in order to obtain data at common rescaled distances and compute statistics on rescaled profiles $s(r')$ (Fig. \ref{fig1_resc}c)).
Since the generic shape of those radial profiles of land use $s_N(r)$ is exponential, we perform an exponential regression for all cities' radial profiles, following $s_N(r)=a_N \exp(-r/l_N)$, where $a_N$ is the share of built-up land in the city center and $l_N$ the characteristic decrease distance, for a city of population $N$. Note that this characteristic distance $l_N$ is defined up to a multiplicative constant. Here it gives the distance at which built-up land is $\exp(-1)\simeq 37\%$ of its value in the center, and also the distance at which the highest quantity of land is built (see Supp. Info.). These regressions give a median R$^2$ of 0.9 on all cities (see Supp. Table 1), a very high value which confirms their quality. We single out waterfront cities and polycentric areas in our analysis. We consider a city as waterfront if more than 60\% of its maximal urban extent disc (of radius $r_{\text{max}}$ around the center) is covered by water bodies, and as interconnected (polycentric) if more than 75\% of its disc overlaps with other cities' maximal extents.
The clear scaling behaviour observed here suggests to define an Urban Built Land Index (UBLI) to capture size-independant variations, as $\text{UBLI}=a_N l_N^2/N$, which is closely related to the residuals of Figure \ref{fig1_resc}d) (as illustrated on Supp. Fig. 11a)). It gives a measure of the characteristic built-up area per inhabitant (defined, as the characteristic distance $l_N$, up to a multiplicative constant), which is on average here 21 m$^2$ (median 12 m$^2$, see Fig. 3), and of a characteristic population density ("net" density, with respect to the built-up area only), on average 47,000 inhabitants/km$^2$ (median 84,000 inhab./km$^2$). 

We note that the analyses are conducted using the R programming language,  v4.2.2. Adding to R-base, R packages are used for spatial analysis (rgdal, rgeos, raster, sp, sf and terra), data manipulation (tidyverse, stringr, purrr, plyr, dplyr, tidyr, data.table, tibble and vctrs), import and export (openxlsx), regressions (minpack.lm, nls2), visualization (ggplot2, plotly, leaflet, mapsf, cowplot, ggthemes, viridis, and ggsn) and more miscellaneous uses (str2str, gdata, corrplot and ggpubr). The processing uses the national shared academic cloud server of the French Very Large Research Infrastructure for Social Sciences and Humanities, Huma-Num.

\backmatter

\bmhead{Supplementary information}

This manuscript has an accompanying supplementary file. 

\bmhead{Acknowledgements}

The authors acknowledge comments by M. Barth\'{e}l\'{e}my, A. Litvine, G. Caruso, R. Le Goix and A. P\'{e}cheric, data provided by P. Kilgarriff as well as funding by ANR GreenLand and RIN SUCHIES.

\bmhead{Author Contributions}
R.L. designed the study. V.V. acquired the data and performed the analyses under supervision by R.L. V.V. and R.L. interpreted the results and wrote the manuscript.

\bmhead{Competing Interests}
The authors declare no competing interests.


\bibliography{biblio}

\appendix

\section*{Supplementary Information}

\subsection*{The GHSL built surface data }
\begin{figure}[h]
\centering
\includegraphics[width=0.98\textwidth]{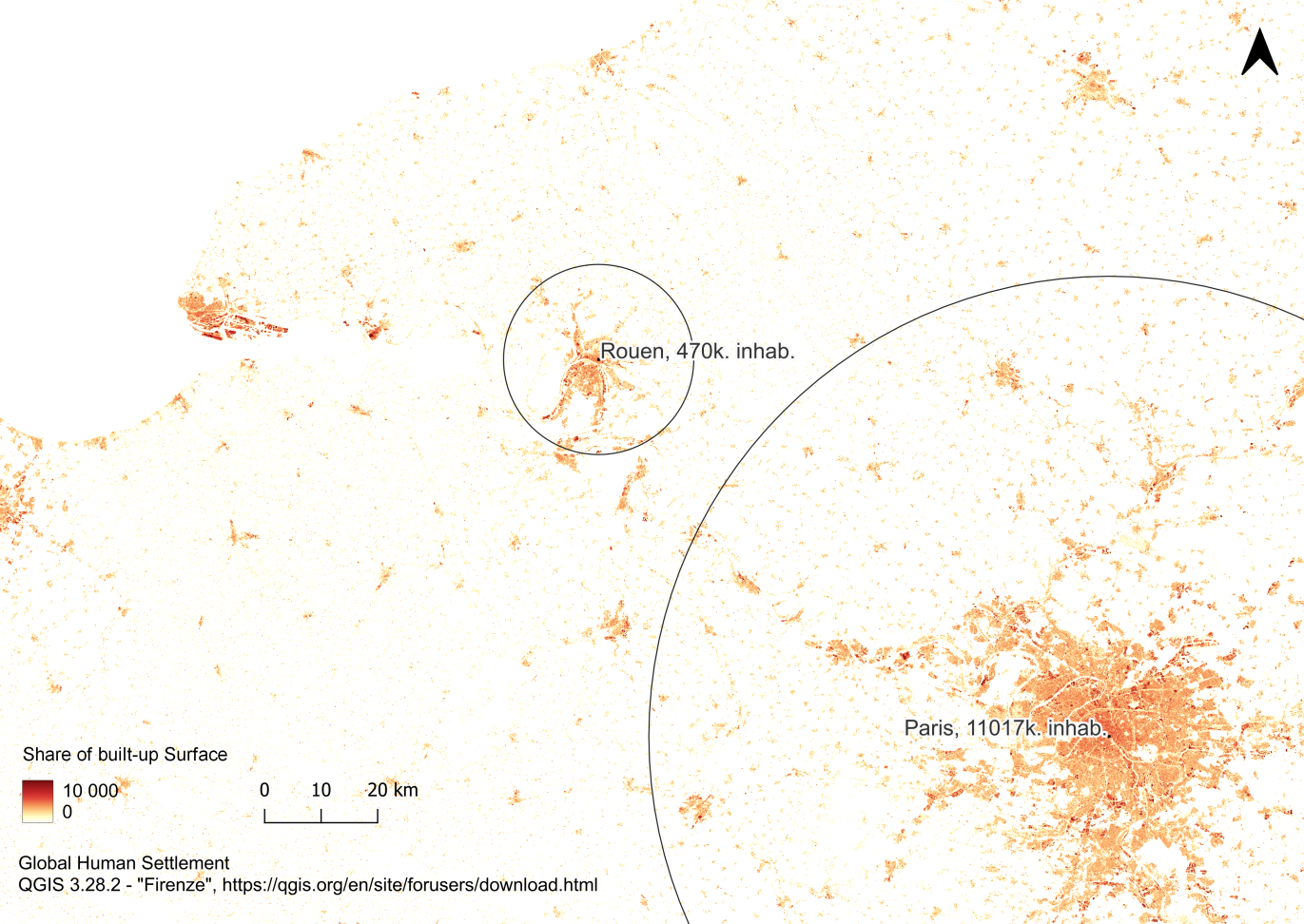}
\caption{The GHSL built surface dataset for the Seine Valley in France (Rouen - Paris axis). A pixel with value 10,000 is fully built, while a pixel with value 0 is not built at all.}\label{fig6_GHSL}
\end{figure}

The main dataset used in this study, the GHS-BUILT-S dataset, is presented here on figures \ref{fig6_GHSL} and \ref{fig7_buffers}. This global raster dataset provides the built-up area (in square meters) within square cells of 100m side length. The value is encoded as a 4 digit integer, between 0 and 10,000 (which corresponds to a 100\% built-up cell). The remote-sensing and processing of this dataset \cite{GHSL2020} ensures that the urban footprints correspond to reality, even in specific areas such as industrial zones, ports or slums. In this way, the entire urban area can be taken into account in this work.

\subsection*{Comparison of urban extents for cities of different sizes}

\begin{figure}[h]
\centering
\includegraphics[width=0.98\textwidth]{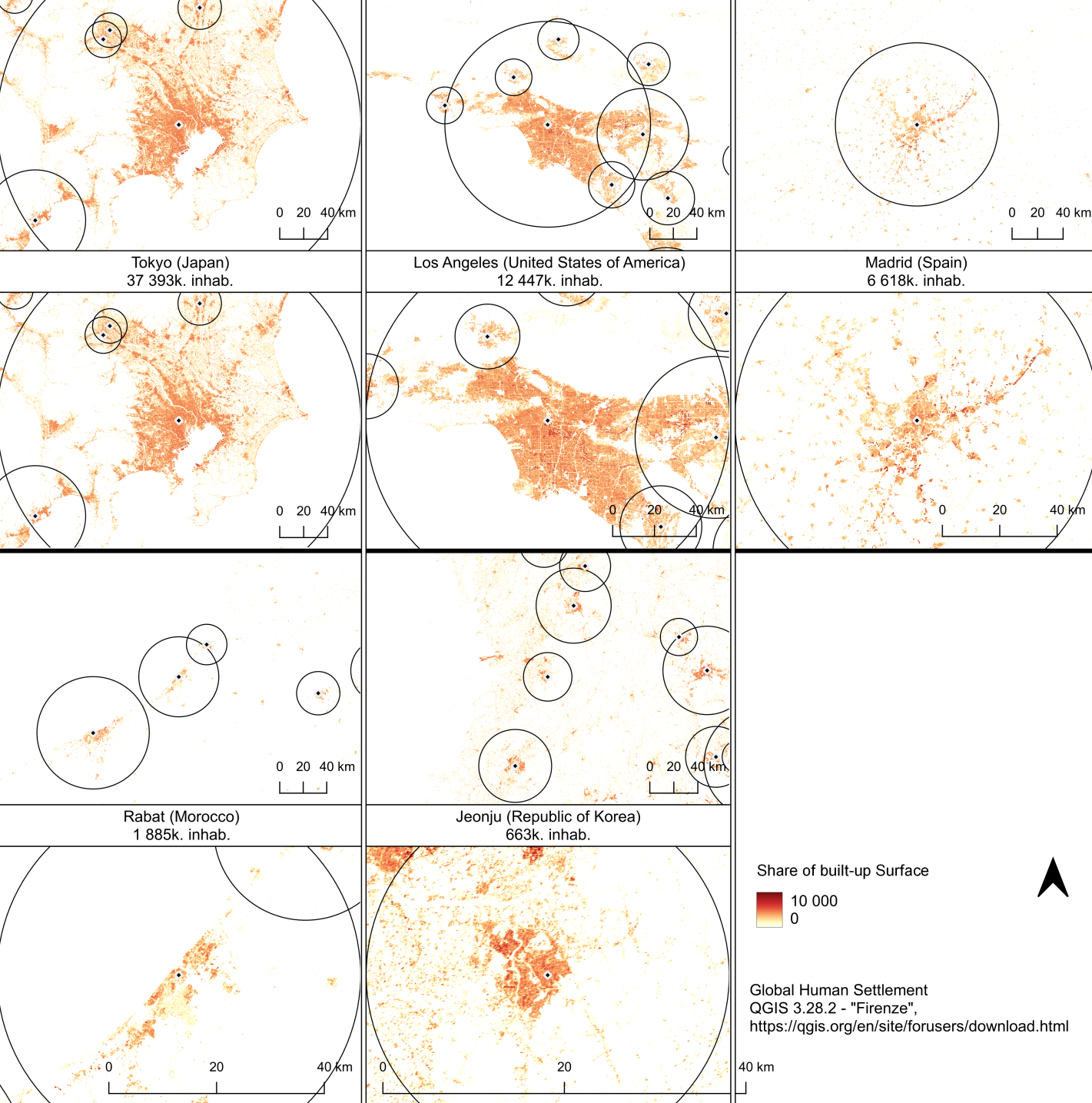}
\caption{Comparison of buffer sizes, with a reference to the population of Tokyo using a $r_{N\text{max}}=$150km radius with a $\alpha=1/2$ scaling exponent. Comparison of buffer sizes for cities of different sizes: Tokyo (Japan), 
Los Angeles (USA), Madrid (Spain), Rabat (Morocco), Jeonju (Republic of Korea). We keep the same scale on all maps of the first and third lines, but the maps of the 2nd and 4th lines are rescaled.
}\label{fig7_buffers}
\end{figure}

We propose to visualize on Figure \ref{fig7_buffers} the maximal urban extents used in this study, in order to understand the chosen definition of cities' buffers. We compare Tokyo's $r_{\text{max}}=$150km radius buffer with its equivalent $r_{N\text{max}}$ in other cities of given population $N$, using the formula $r_{N\text{max}}=r_{\text{max}}/k_N=r_{\text{max}} \sqrt{N/N_{\text{Tokyo}}}$. 

This method allows us to circumvent one major problem in the study of the urban environment, which concerns the definition of the city. There is no consensus on how they should be defined, either by researchers or by other organizations. And this clearly influences the results of comparative urban studies \cite{louf2014}. Taking or not taking into account specific zones will lead to different results, since perimeters are not analogous. There is a lack of a uniform definition of cities across the world, which makes it difficult to compare urban areas \cite{seto2011}. 
In this work, we manage to use this global database  because our buffer definition is based on the population of cities, which is defined here by the World Urbanization Prospects of the United Nations. We note that in this population dataset each year's population includes collected data (national censuses, sample surveys, civil registers, etc.), and demographic projections to make up for the fact that harvesting dates can vary from one country to another, and for data gaps.

\subsection*{Signal over noise ratio}

\begin{figure}[h]
\centering
\includegraphics[width=0.6\textwidth]{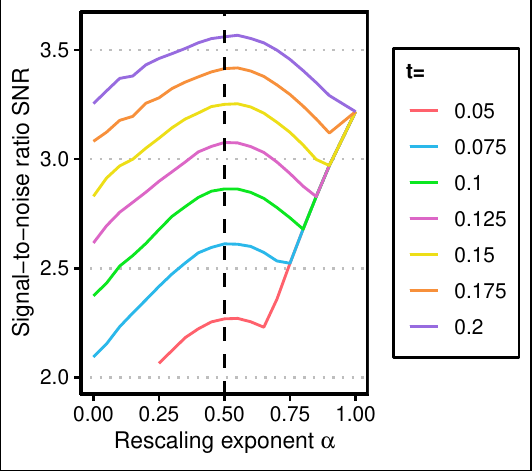}
\caption{Variation of the signal to noise ratio SNR as a function of the rescaling exponent $\alpha$ used, for different threshold values $t$. The black vertical line represents exponent $\alpha=1/2$. The linear behavior for large values of $\alpha$ is linked to the maximal urban extent $r_{\text{max}}$ being reached.}\label{SNR}
\end{figure}

A signal-to-noise ratio SNR can be computed on the (rescaled) built-up radial profiles $s_N(r')$ to determine the most appropriate scaling exponent \cite{lemoy2020}. To compute this quantity, we rescale the profiles with different exponent values $\alpha$ ranging from 0 to 1, meaning that the rescaled distance is $r'=r(N_{\text{Tokyo}}/N)^{\alpha}$. Then we compute a measure of the efficiency of this rescaling. We start with the  mean value $<s_N(r')>$ (over all cities) of the built-up land share at each (rescaled) distance $r'$, which we consider as our signal. The corresponding standard deviation $\sigma_s(r')=\sqrt{<(s_N(r')-<s_N(r')>)^2>}$ is the noise. We compute the SNR as the ratio of both quantities, averaged over all distances $SNR=\overline{<s_N(r')>/\sigma_s(r')}$. This averaging over rescaled distances (which we denote with an overline) starts in the center ($r'=r=0$) and stops when the mean built-up land share $<s_N(r')>$ reaches a given threshold $t$ (for instance, $t=0.1$ or 0.175). We then represent the variations of this SNR as a function of the values of the rescaling exponent $\alpha$ and the threshold on Figure \ref{SNR}, which helps us find the best rescaling exponent. Indeed, the highest point on the curves corresponds to the strongest signal over noise ratio and therefore to the most relevant value, for each threshold. Here it corresponds to the scaling exponent 1/2.

\subsection*{Impact of considering water bodies}

\begin{figure}[h]
\centering
\begin{tabular}{cc}
\includegraphics[width=0.45\textwidth]{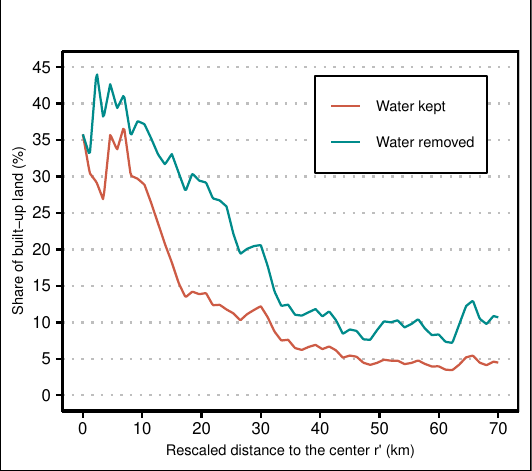} &
\includegraphics[width=0.45\textwidth]{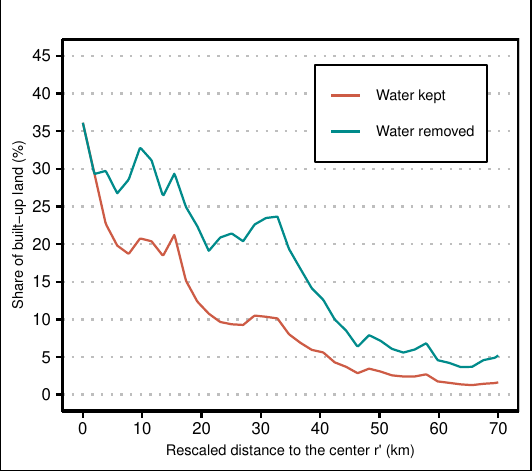}
\end{tabular}
\includegraphics[width=0.5\textwidth]{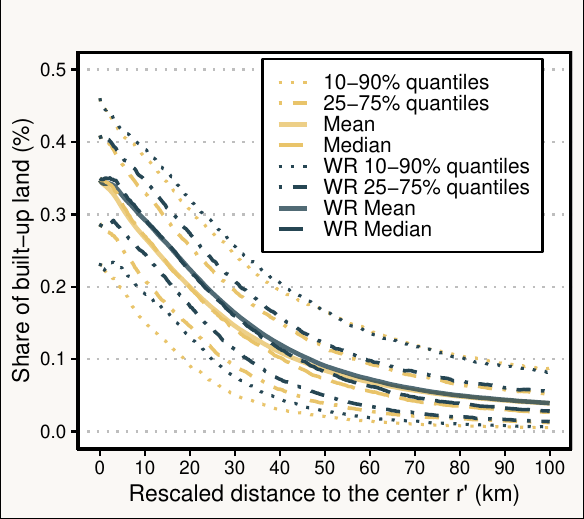}
\caption{Effect of considering water on built-up land shares. Top: for two cities presenting more than 60\% of water bodies within their maximal extent, Da Nang (Vietnam, left) and Las Palmas de Gran Canaria (Spain, right). Bottom: statistics on the studied 1800+ rescaled radial profiles of urban built-up land use. The graph compares profiles where water bodies are kept in the analysis, and where water bodies are removed (WR) from the analysis, with consequently slightly higher shares of built-up land.}\label{fig8_water}
\end{figure}

Some cities have a large amount of water bodies around them. This clearly has an impact on their built-up area in the sense that the presence of water is a constraint for building construction and artificialization. In order to illustrate this, we show on the first line of Figure \ref{fig8_water} how water bodies impact built-up land profiles for two coastal cities. When water is considered, that is, land covered by water bodies is removed from our analysis, built-up land shares are higher, since they are computed on incomplete rings. On the second line, we show that this does not affect the scaling law and the general picture much, since not all cities are coastal.

\subsection*{Characteristic decrease distance and models}

We perform non-linear regressions of the built-up land share $s_N(r)$ as a function of the distance $r$ to the center, following $s_N(r)=a_N \exp(-r/l_N)$. The first parameter $a_N$ gives a measure of the built-up land share in the city center. We note here that the second parameter, the characteristic distance $l_N$, is the distance at which built-up land is at $\exp(-1)\simeq 37\%$ of its value in the center. It is also the distance from the center at which the highest (absolute) amount of land is built, according to our exponential model. Indeed, the total length of built-up land in each ring (considered here of infinitesimal width) is $2\pi r s_N(r)$, which is maximum at $l_N$.

In Table \ref{models}, we present the results of the estimated scaling law $l_N\sim l_1 N^{\alpha}$, using log-log fits on the characteristic distance $l_N$ against population $N$. We either keep or remove water bodies from the analysis, use one- or two-parameter models and keep or remove very polycentric urban areas from the analysis. The one-parameter model uses $a_N=35\%$ and fits only $l_N$. This is associated with only a very small loss in median R$^2$, which shows that this hypothesis is very reasonable. Cities are considered to be in very polycentric urban areas here if the overlap rate of their buffer by other cities' buffers exceeds 75\% (113 cities are in this situation).

We can observe on Table \ref{models} that the fitted scaling exponent $\alpha$ is very close to 1/2, the constant $l_1$ around 6m, the median R$^2$ over all profiles around 0.9 and the scaling law's R$^2$ around 0.45 in most cases (slightly higher for one-parameter models).

\begin{table}
\begin{tabular}{lcccccccc}
Model & NL1$_w$ & NL2$_w$ & NL1*$_w$ & NL2*$_w$ & NL1 & NL2 & NL1* & NL2* \\
\hline
Observations & 1860 & 1860 & 1747 & 1747 & 1860 & 1860 & 1747 & 1747 \\
Scaling exponent $\alpha$ & 0.480 & 0.484 & 0.484 & 0.506 & 0.526 & 0.526 & 0.531 & 0.520 \\
Constant $l_1$ (m) & 7.38 & 7.44 & 6.70 & 5.08  & 4.52 & 4.74 & 4.10 & 4.70 \\
Median R$^2$ (all profiles) & 0.88 & 0.89 & 0.89 & 0.91  & 0.88 & 0.93 & 0.88 & 0.93 \\
Scaling law's R$^2$ & 0.41 & 0.22 & 0.45 & 0.48  & 0.46 & 0.19 & 0.50 & 0.47 \\
\hline
\end{tabular}
\caption{Results of a (log-log) linear regression on characteristic distance $l_N$ against city population $N$, $\log l_N \sim \log l_1 + \alpha \log N $ for different models and samples. Rows: the number of observations is the number of cities considered in the model. The scaling exponent is denoted $\alpha$. The constant $l_1$ is given in meters. The two measured R$^2$ correspond to the median R$^2$  of the corresponding non-linear fit over all profiles, and the R$^2$ of our (log-log) linear regression of the scaling law. Columns: each non-linear model processed on the 1800+ cities with one parameter (NL1, where $a_N$ is fixed to 35\%) and two parameters (NL2), and on a sample of 1747 cities with one and two parameters (NL1* and NL2*). In the first 4 models, which have a $w$ index in their names, water is not considered (that is, water bodies are kept in the computation of radial profiles).}
\label{models}
\end{table}
\backmatter

\subsection*{Distribution of the 1860 studied cities, and of peculiar characteristics: water bodies and polycentric regions}

\begin{figure}[h]
\centering
\includegraphics[width=0.98\textwidth]{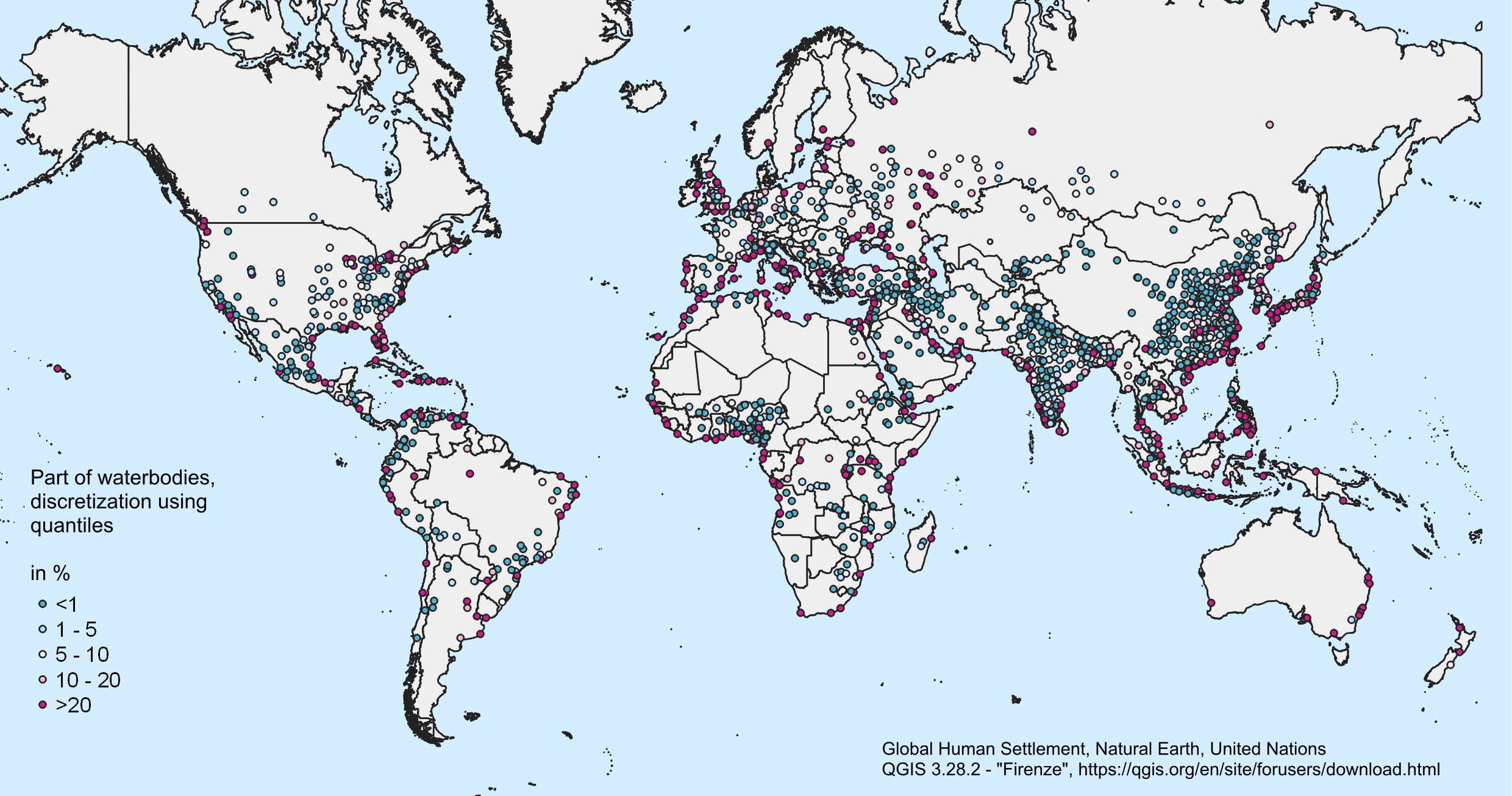}

\includegraphics[width=0.98\textwidth]{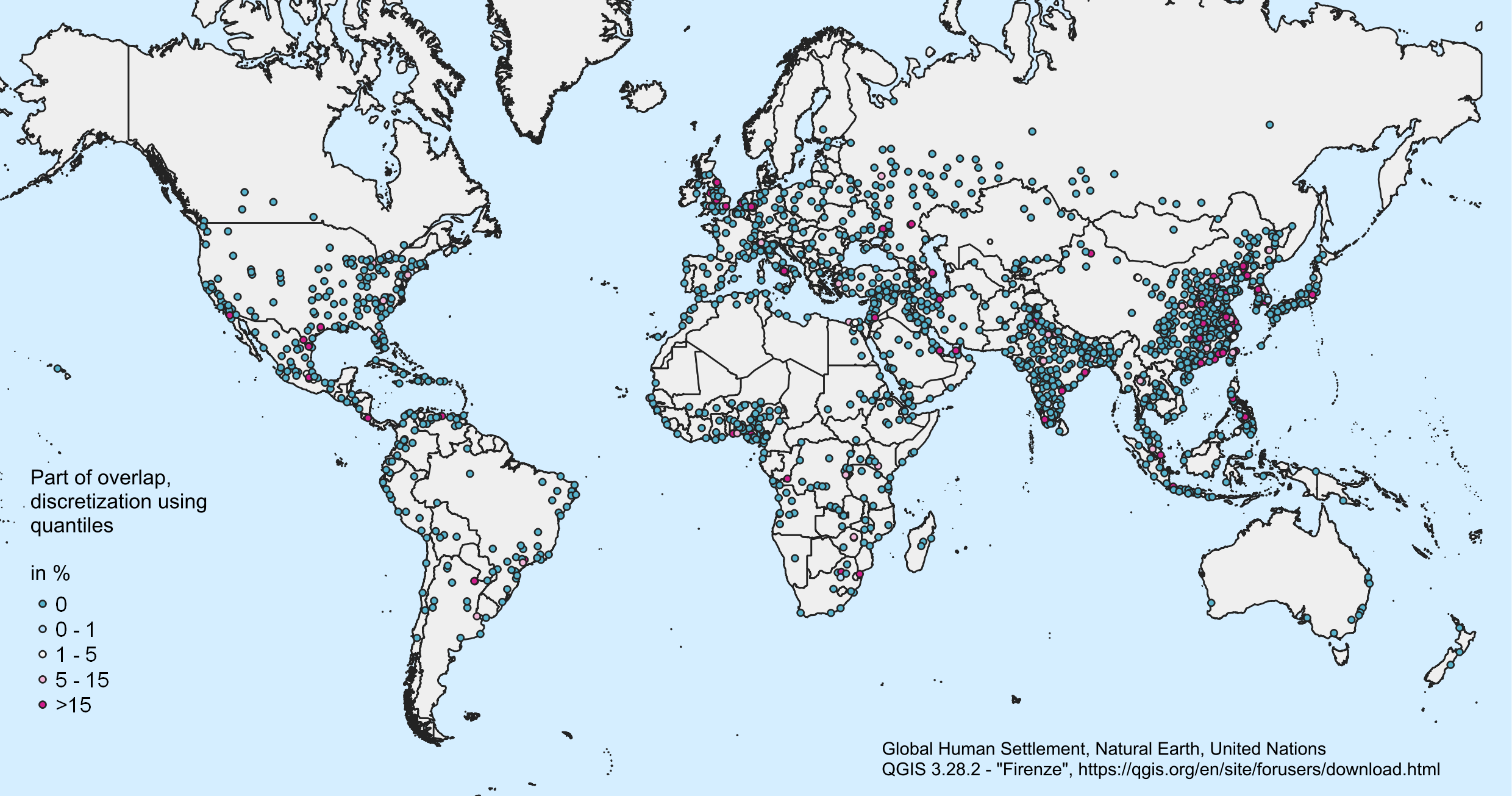}
\caption{Geographical distribution of urban areas with 300,000 or more inhabitants in 2020. We identify two main types of determining characteristics: top panel, the presence of an important water body, bottom panel, the proximity with other cities. }\label{fig5_distri}
\end{figure}

We identify two important factors influencing urban land use, whose variations at the global scale are represented on Figure \ref{fig5_distri}. The first one is the presence of water bodies, which is mostly associated with coastal cities. And the second one is the existenc of polycentric urban regions, which we measure by the overlap rate of a city's maximal extent buffer of radius $r_{N\text{max}}$ with other cities' extents. This is associated with the most populated areas of the world.

\begin{figure}[h]
\centering
\begin{tabular}{cc}
\includegraphics[width=0.45\textwidth]{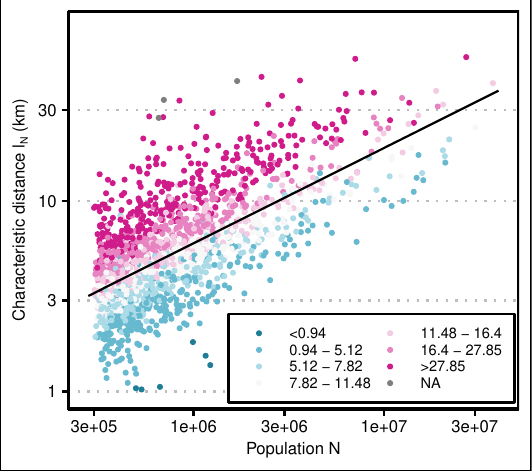} &
\includegraphics[width=0.45\textwidth]{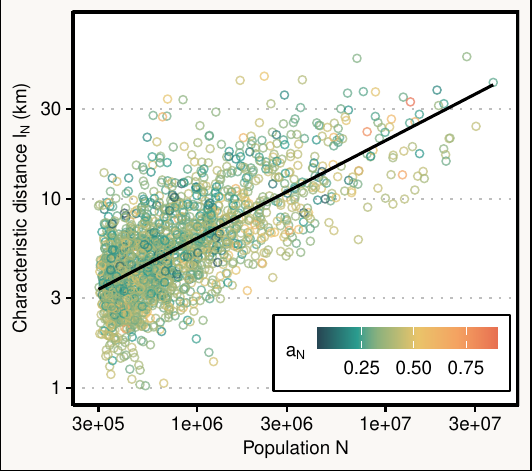}
\end{tabular}
\caption{Characteristic distance computed with the NL2* model. The colors indicate the Urban Built Land Index UBLI$=a_N l_N^2/N$ in m$^2$/inhabitant (left panel), which is directly related to the residual of the regression, and the share of built-up land in the center $a_N$ (right panel). On the left panel, the line is $l_N=l_1 \sqrt{N}$, where $l_1=6$ m, while on the right panel it is $l_N=0.0047 N^{0.52}$.}\label{lN}
\end{figure}

\begin{figure}[h]
\centering
\includegraphics[width=0.98\textwidth]{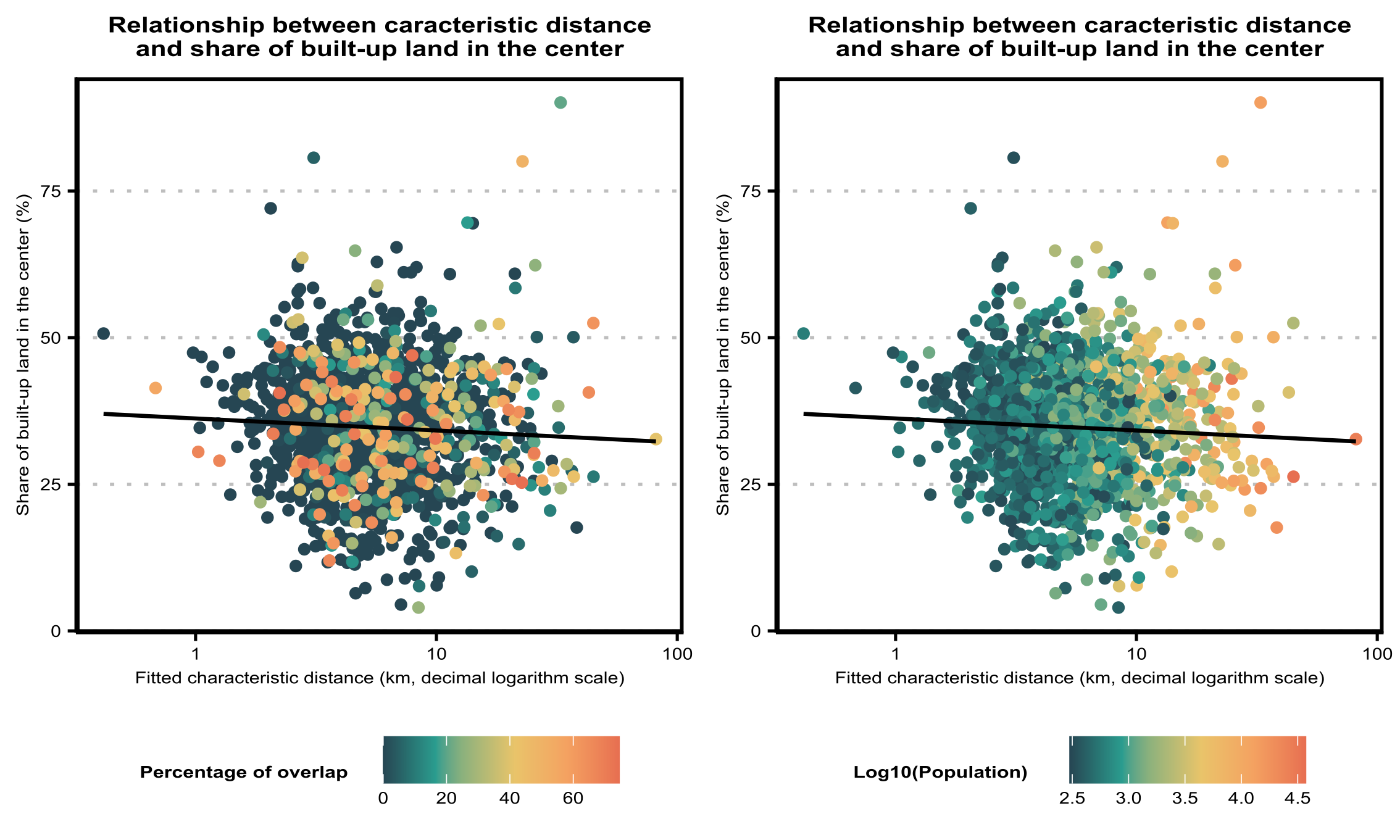}
\caption{Relationship between the share of built-up surface in the center $a_N$ and the characteristic distance $l_N$ computed with the SNL2* model. The colors indicate the percentage of overlap with neighboring cities (left panel) and the total population $N$ (right panel). }\label{a_N}
\end{figure}

\begin{figure}[h]
\centering
\includegraphics[width=0.98\textwidth]{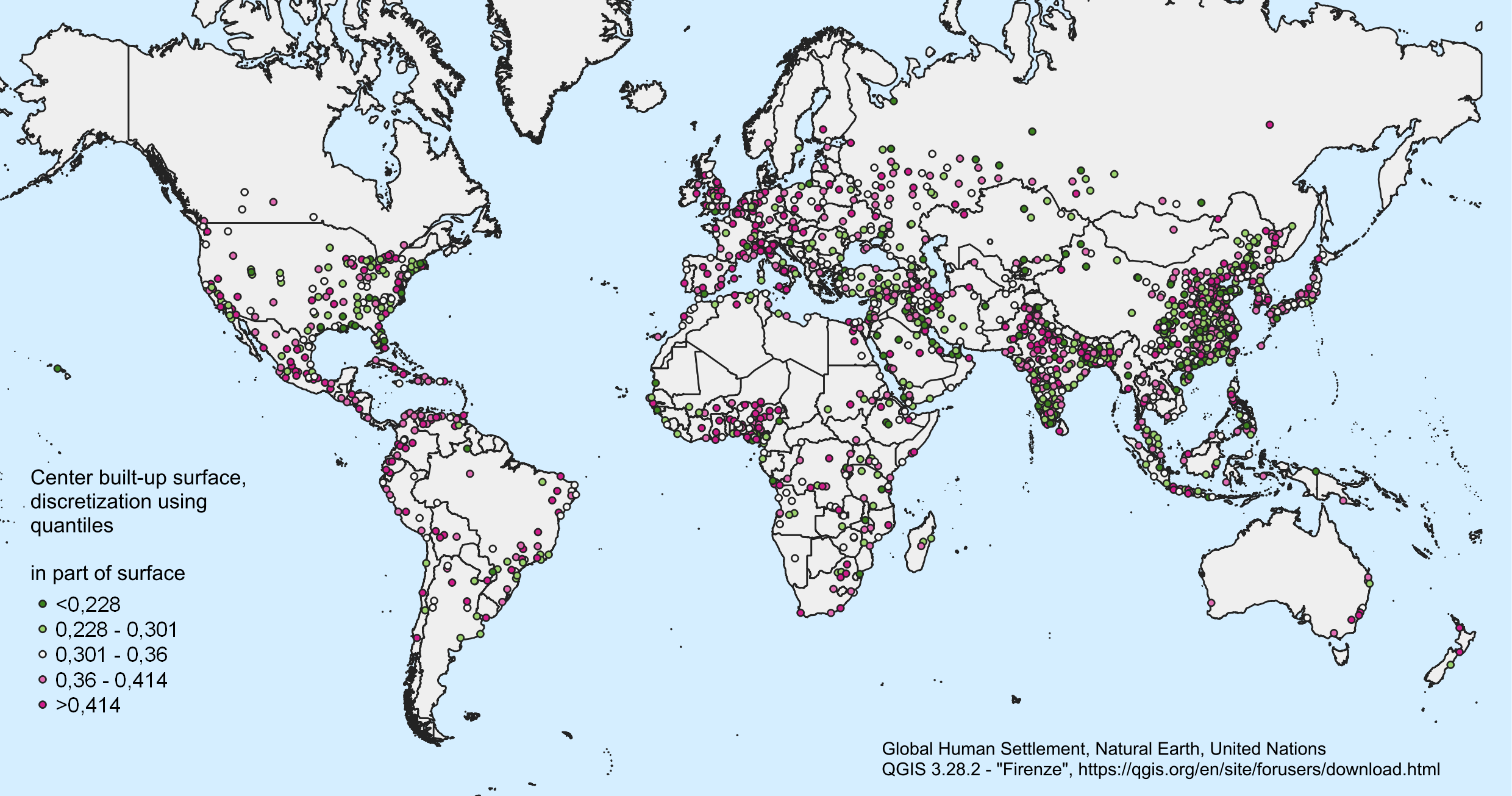}
\caption{Built-up surface in the center $a_N$ for the 1800+ cities.}\label{aN}
\end{figure}

\subsection*{Urban Built Land Index UBLI}
We note here that the Urban Built Land Index defined in the main text UBLI$=a_N l_N^2/N$ is very closely linked to the residual of the power-law relationship presented on Fig. \ref{lN} (and Fig. 1d)). Indeed this relation can be written as $l_N=l_1 \sqrt{N}$, or $\log l_N = \log l_1 + 1/2 \log N$. The residual of this relation can be written as $\log l_N - (\log l_1 + 1/2 \log N)$, while $1/2 \log \text{UBLI} = 1/2 \log a_N + \log l_N - 1/2 \log N$. The difference between both quantities is then only $\log l_1$, which is a constant, and $1/2 \log a_N$, which is almost constant, as seen before.

We show on Figures \ref{lN}, \ref{a_N} and \ref{aN} that the fitted share of built-up land in the center $a_N$ is indeed roughly constant, around 35\%. It is quite independant of the characteristic distance $l_N$ (and hence population size $N$), and does not show a clear relationship with polycentric urban areas either.

\begin{figure}[h]
\centering
\includegraphics[width=0.6\textwidth]{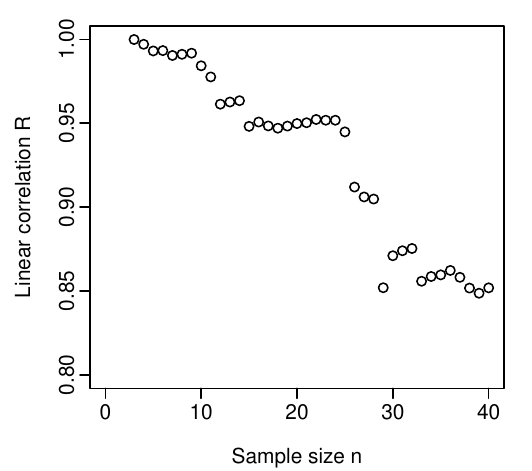}
\caption{Variation of the correlation $R$ between median UBLI and GDP pc as a function of the number $n$ of countries included in the analysis. Countries are included in the sample in order of decreasing number of cities.}\label{corr}
\end{figure}

\begin{figure}[h]
\centering
\includegraphics[width=0.98\textwidth]{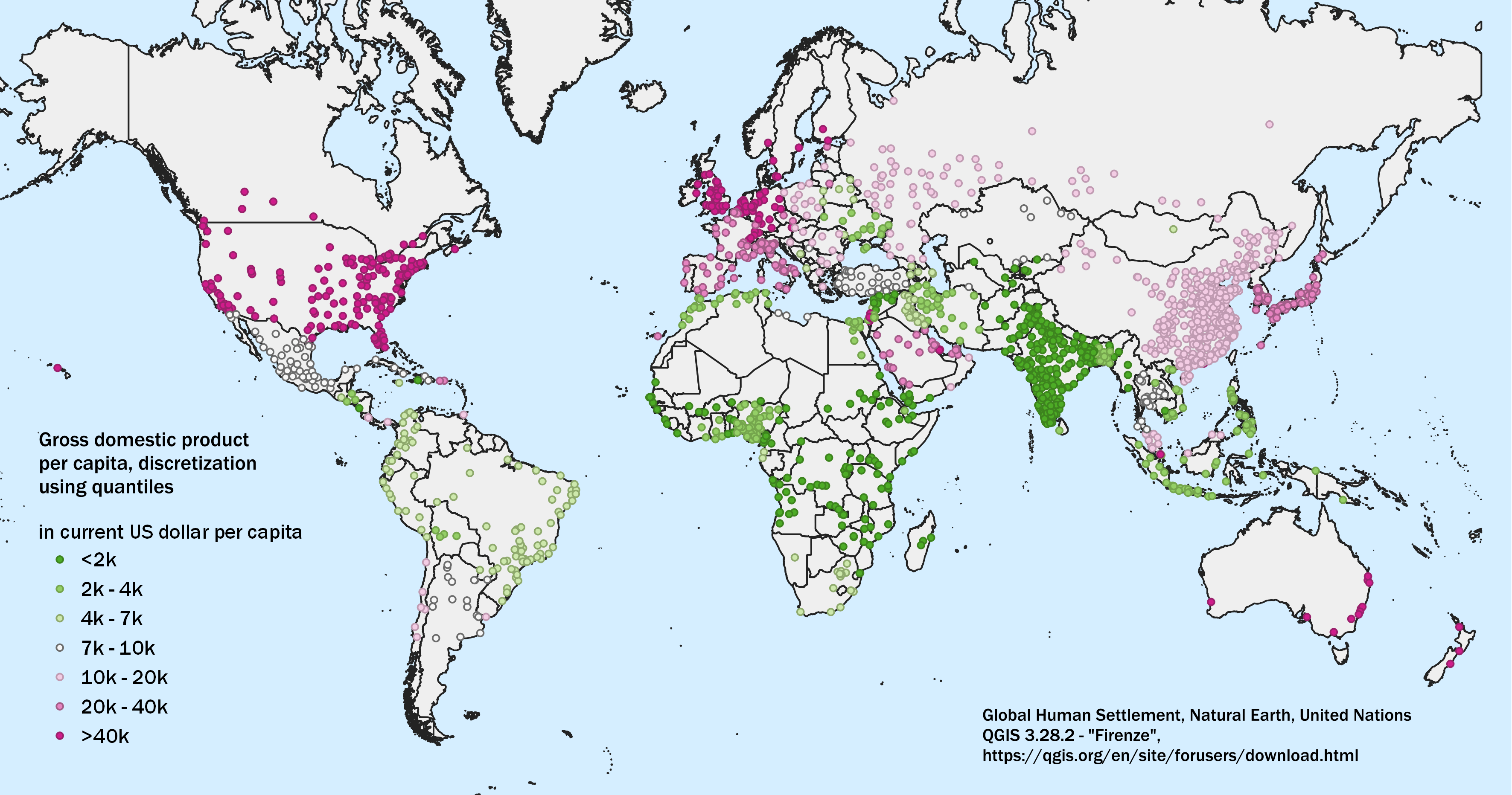}
\caption{Map of the national Gross Domestic Product GDP per capita, where the colors are applied to the studied cities, for comparison with the map of the UBLI (Fig. 3 of the main text).}\label{GDPmap}
\end{figure}

\begin{figure}[h]
\centering
\includegraphics[width=0.92\textwidth]{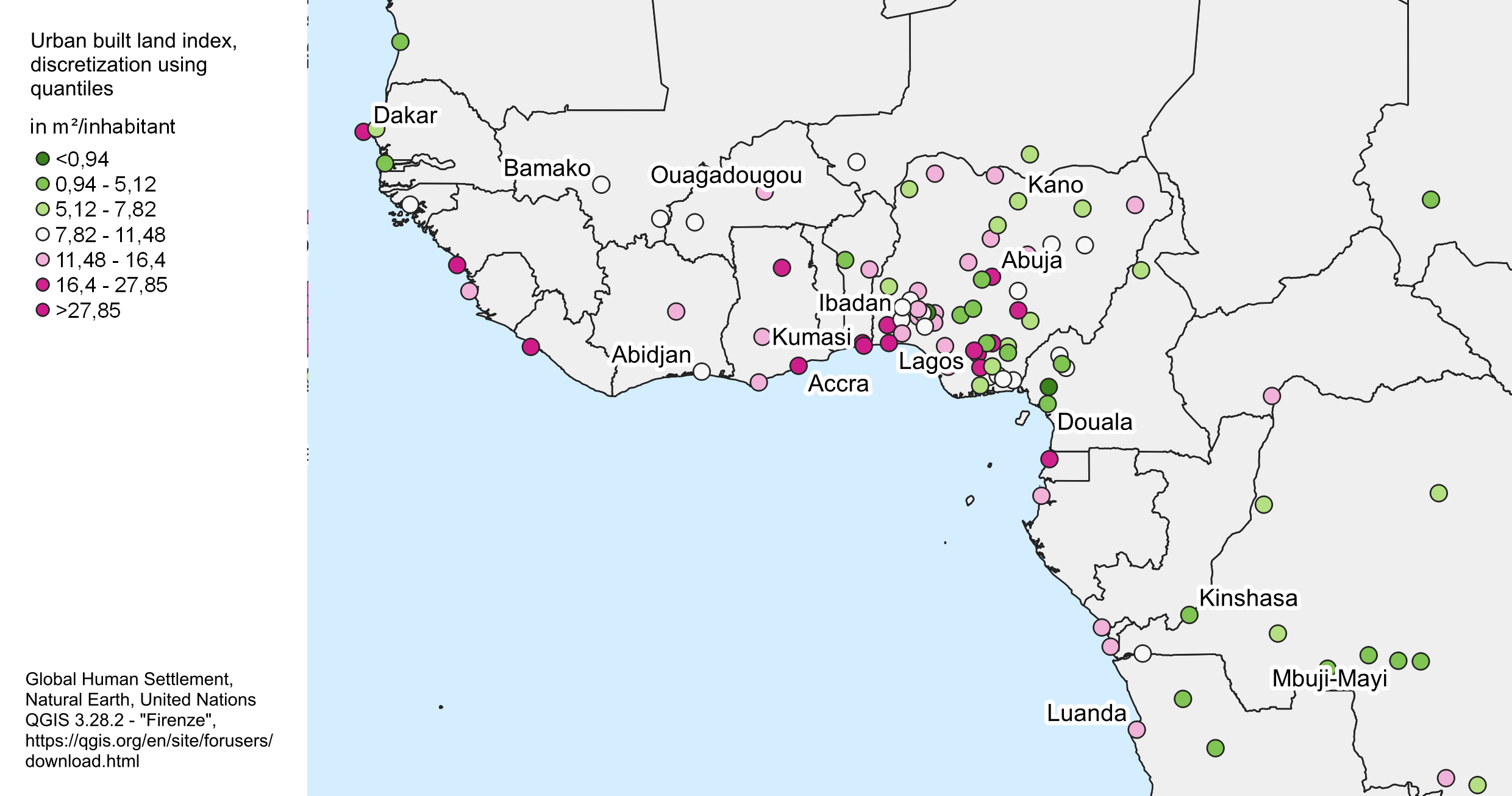}
\includegraphics[width=0.92\textwidth]{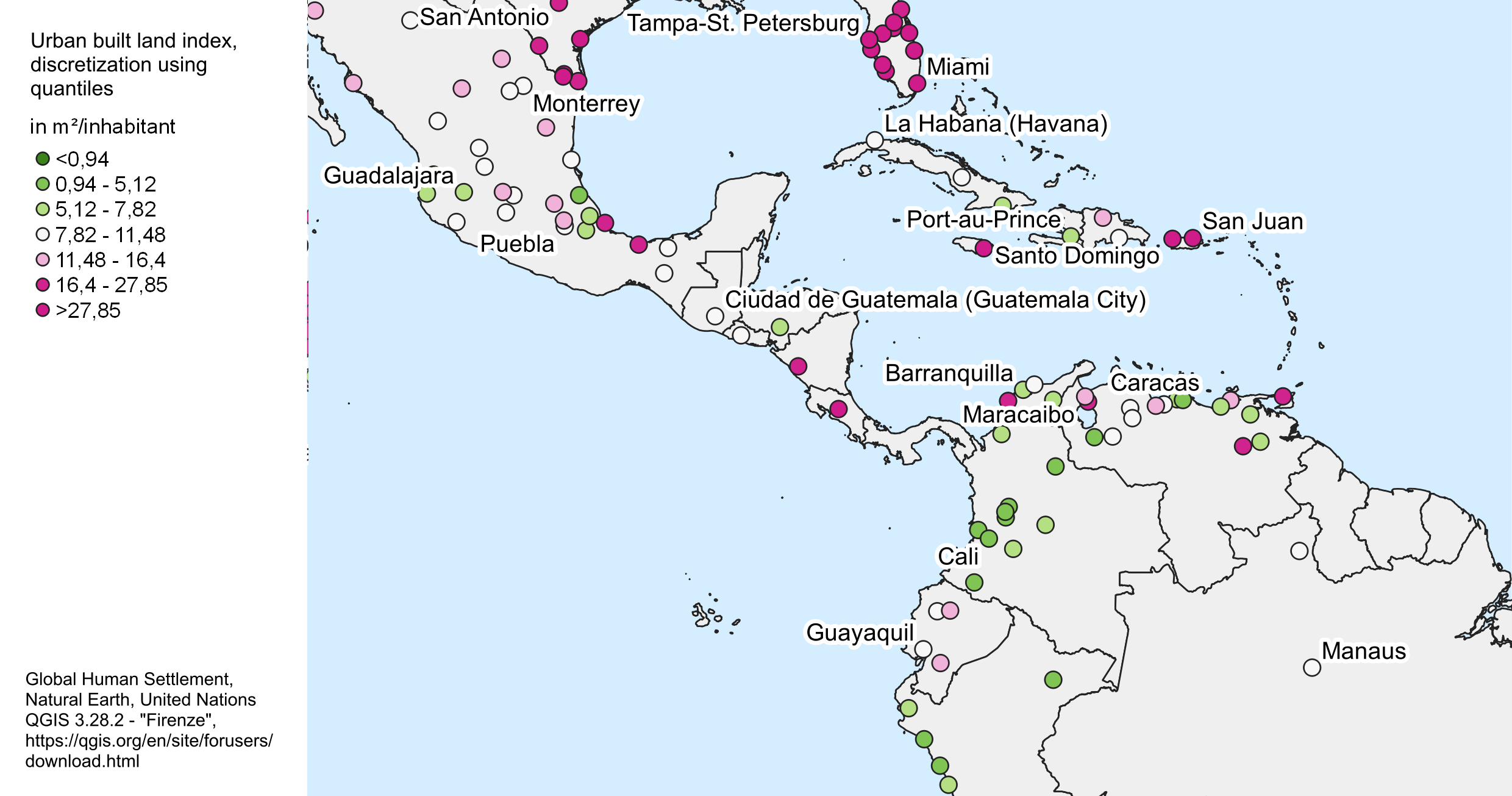}
\includegraphics[width=0.92\textwidth]{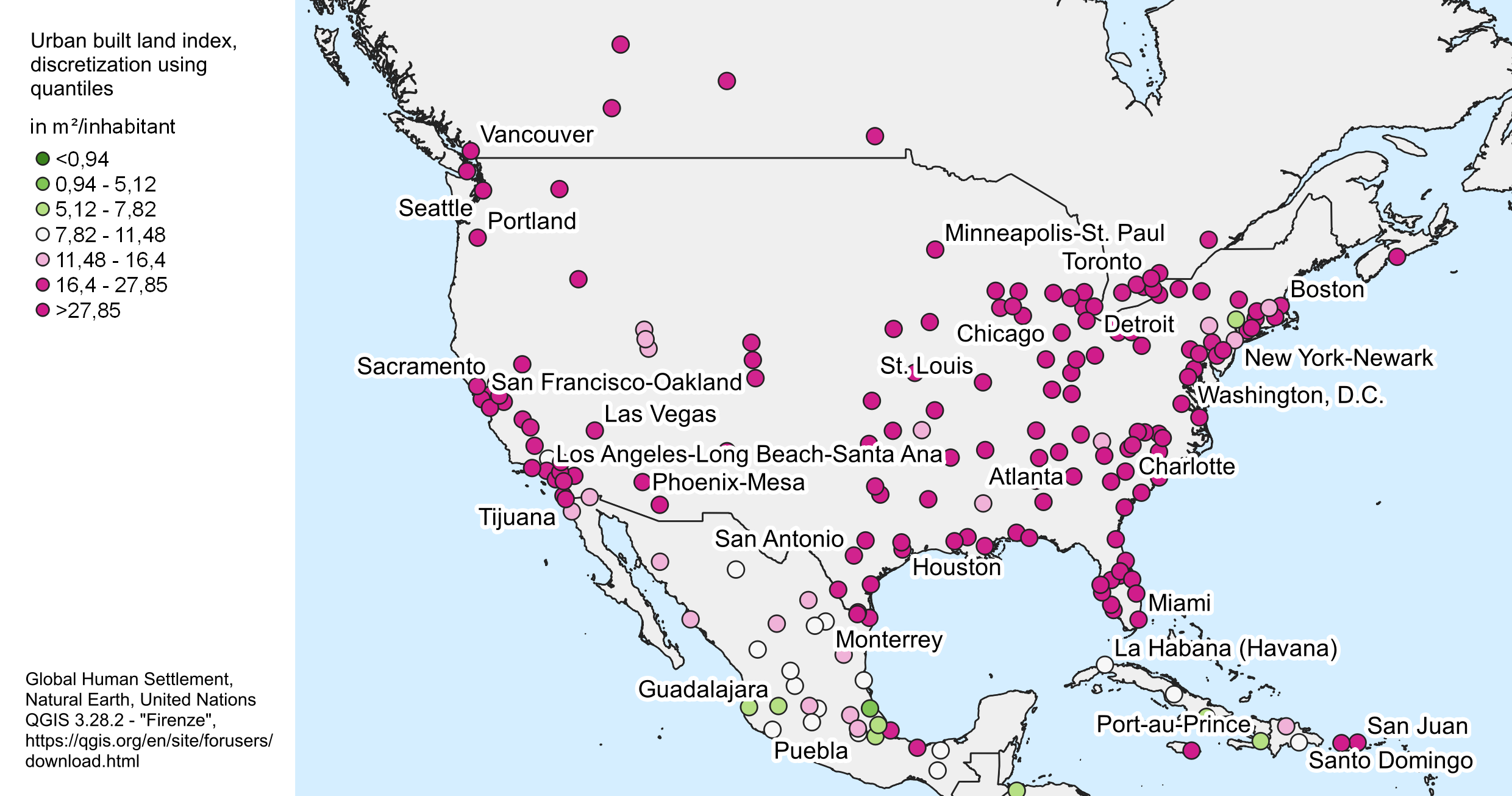}
\caption{Zooming on different parts of the global map of the Urban Built Land Index UBLI (Fig. 3 of the main text): West Africa, Central America, North America.}\label{zooms1}
\end{figure}

\begin{figure}[h]
\centering
\includegraphics[width=0.92\textwidth]{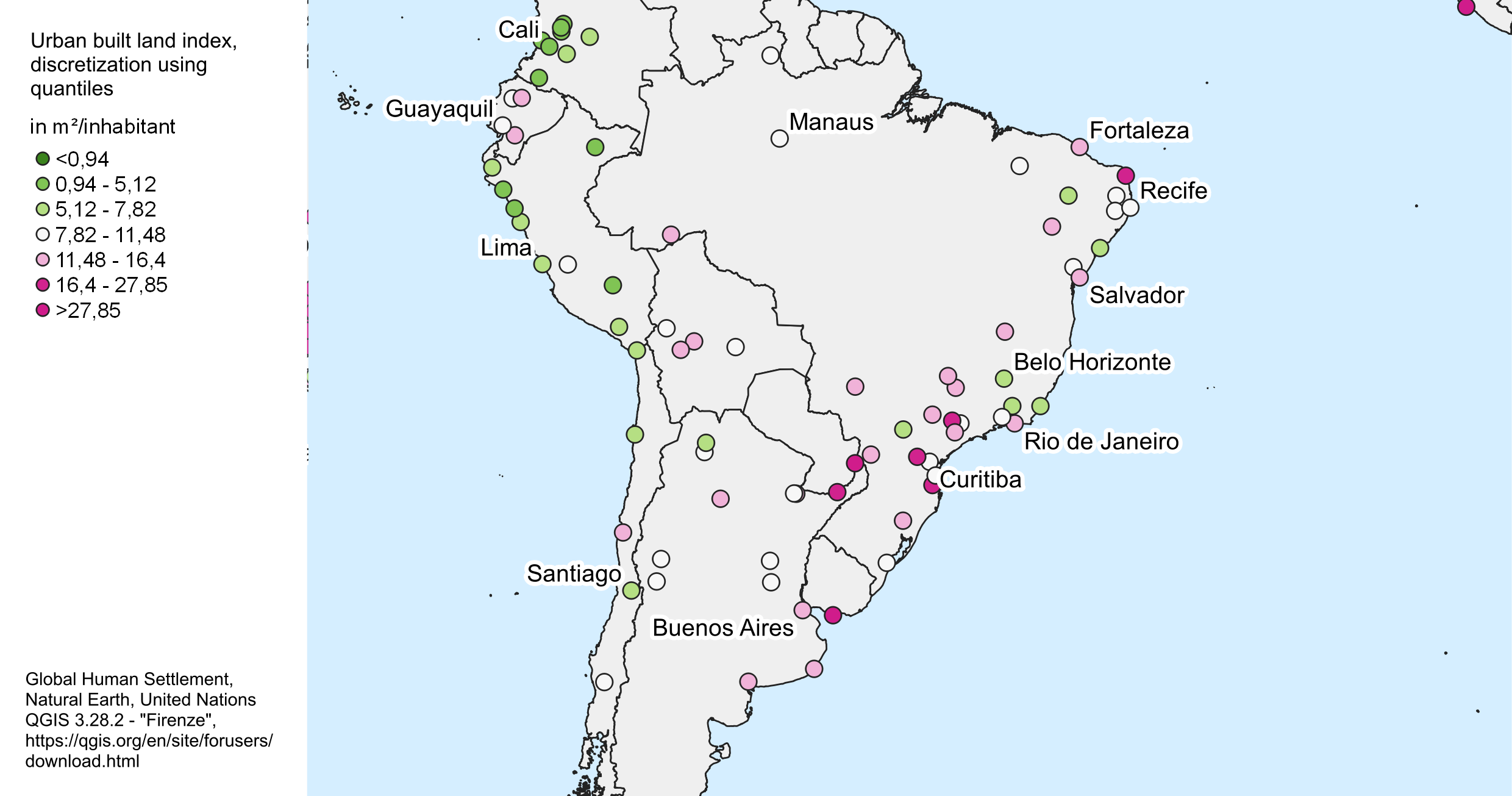}
\includegraphics[width=0.92\textwidth]{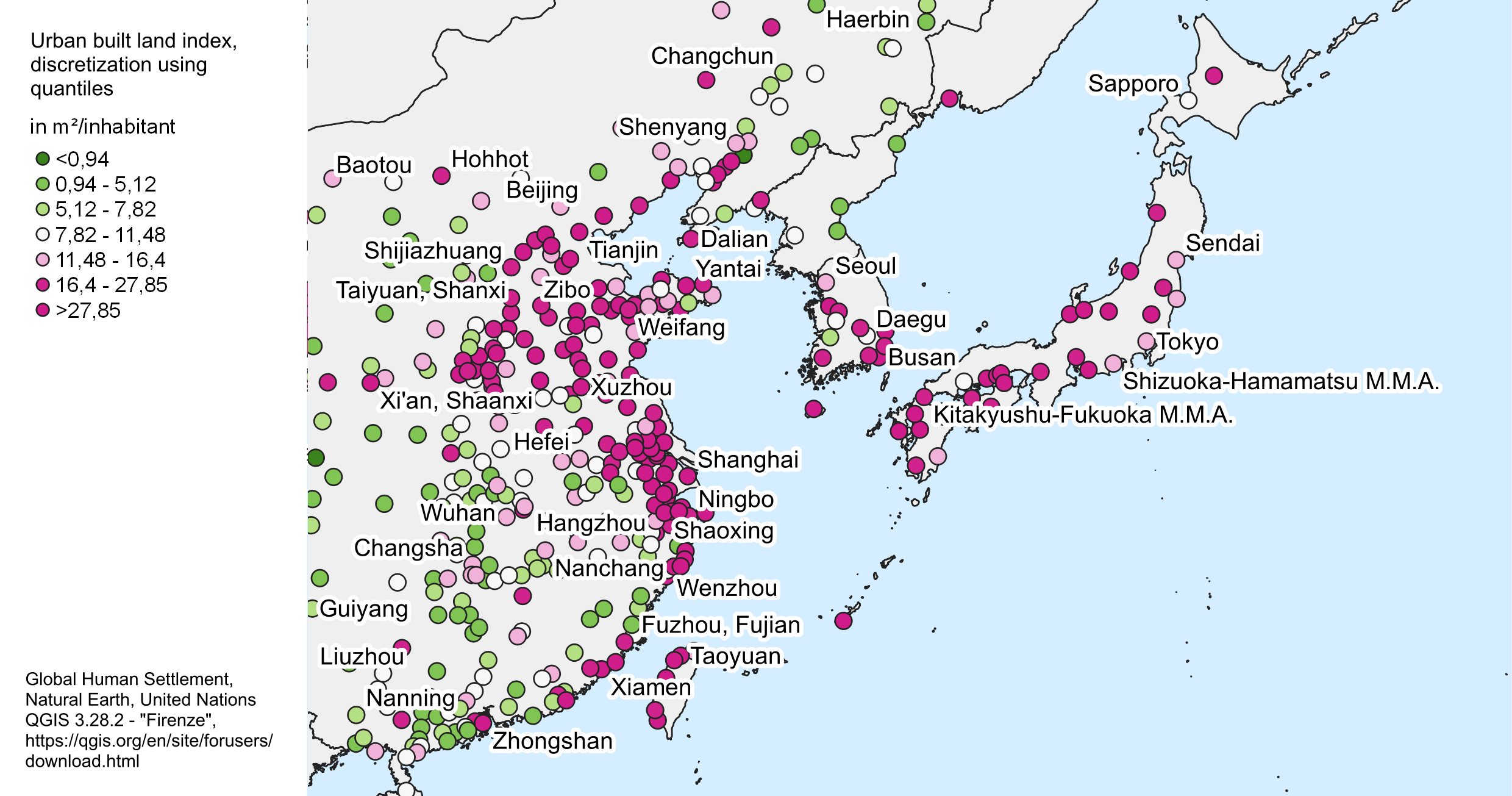}
\includegraphics[width=0.92\textwidth]{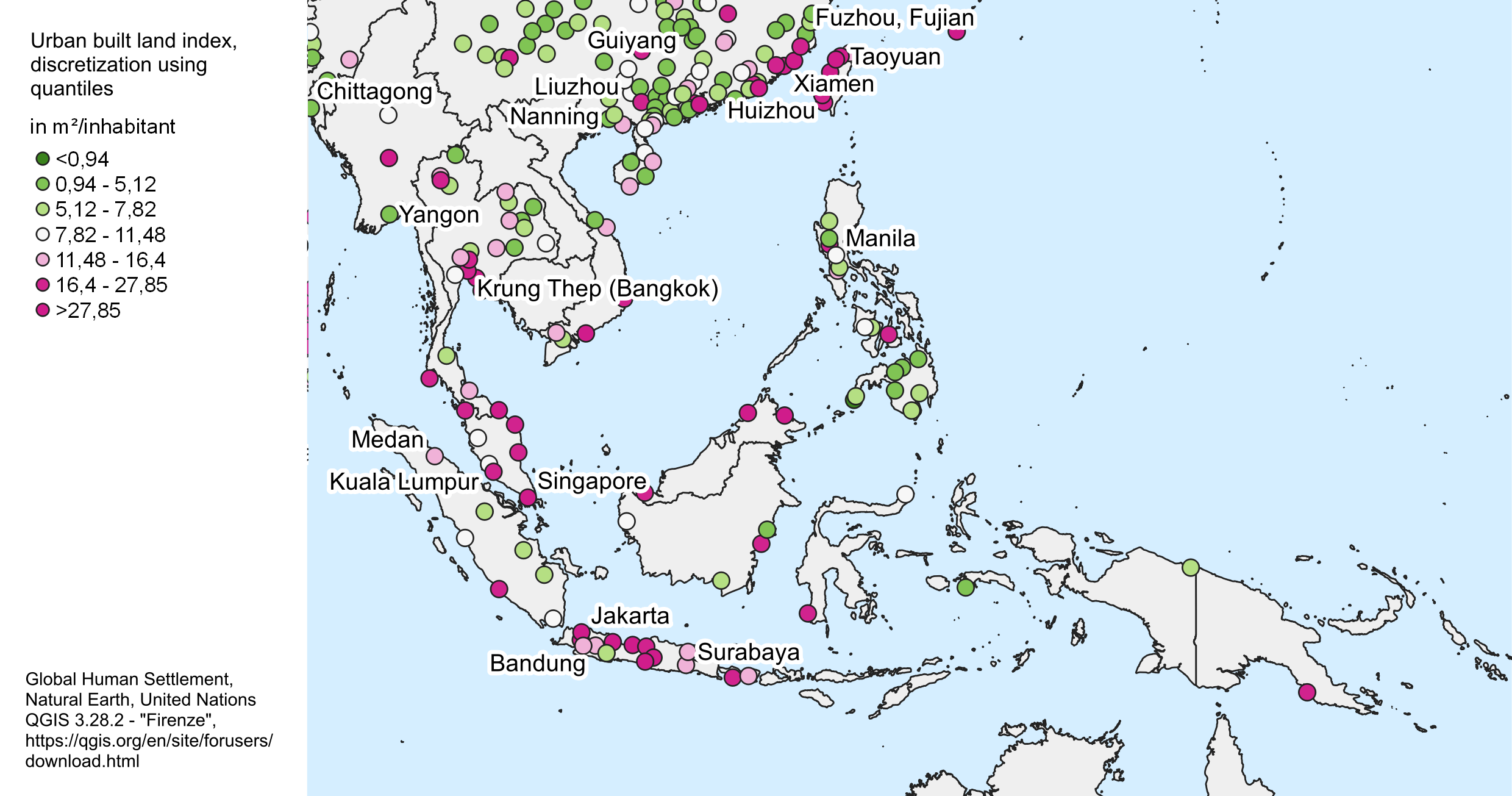}
\caption{Zooming on different parts of the global map of the Urban Built Land Index UBLI (Fig. 3 of the main text): South America, East Asia, SouthEast Asia.}\label{zooms2}
\end{figure}

\begin{figure}[h]
\centering
\includegraphics[width=0.92\textwidth]{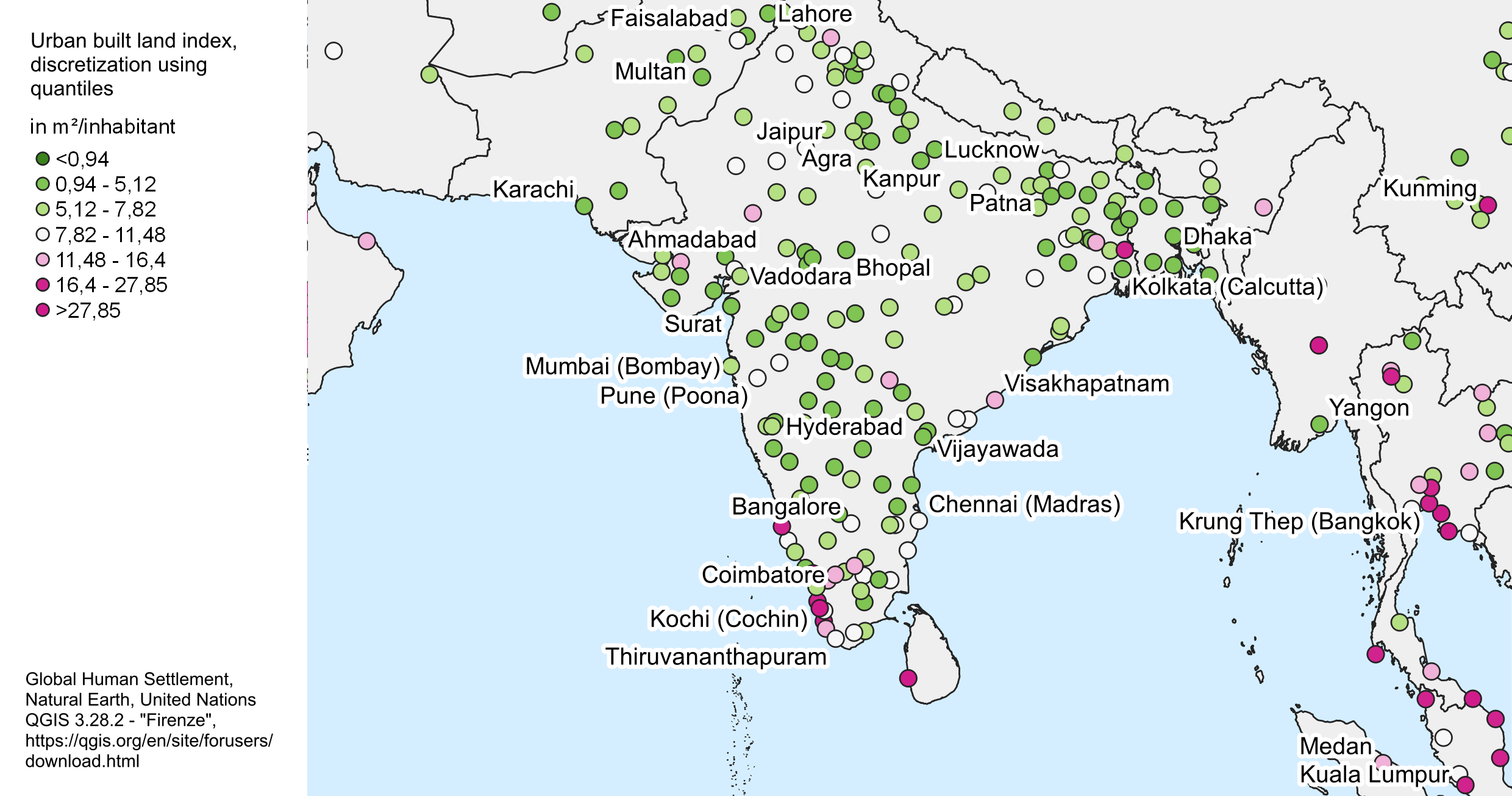}
\includegraphics[width=0.92\textwidth]{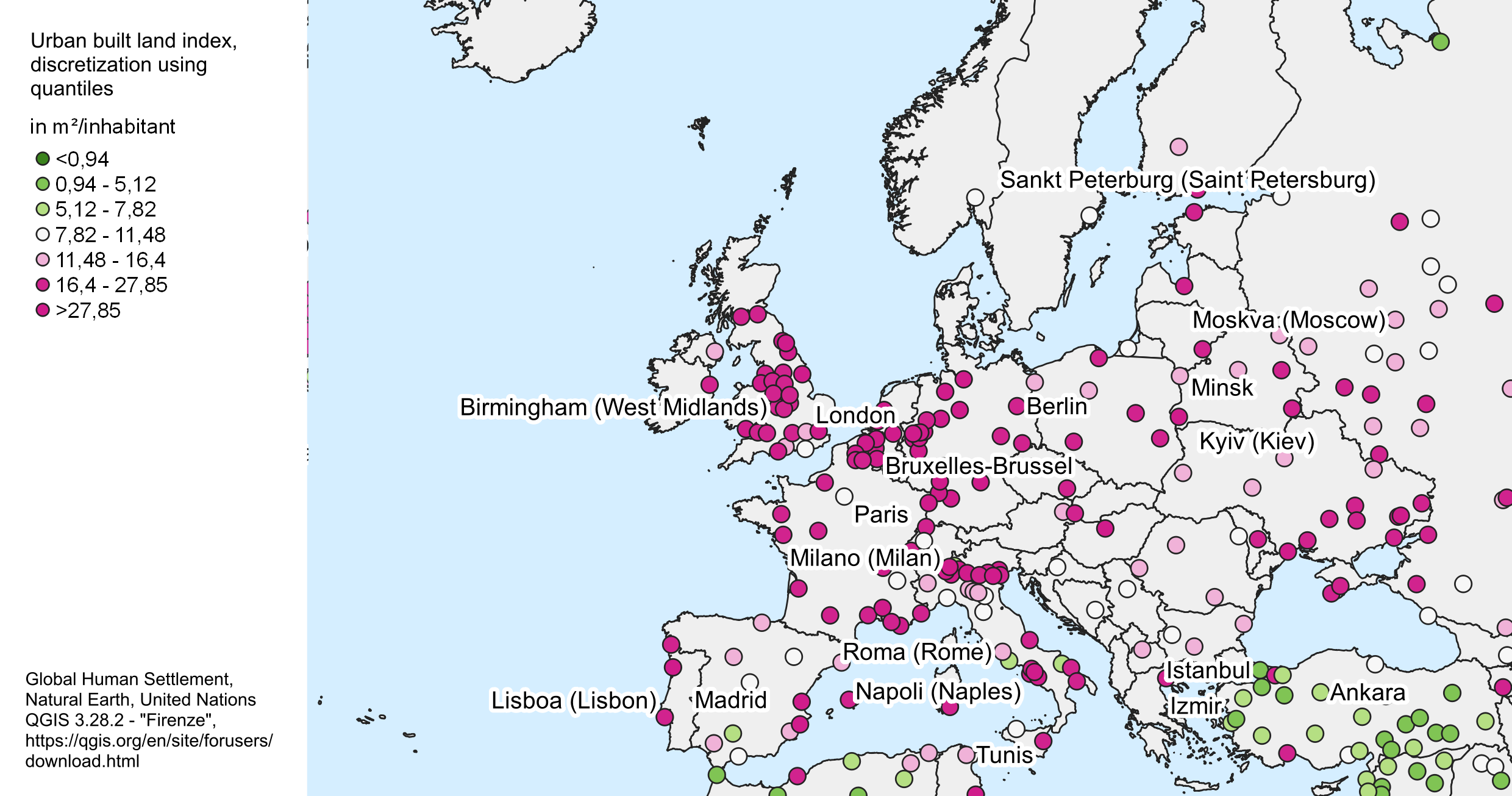}
\includegraphics[width=0.92\textwidth]{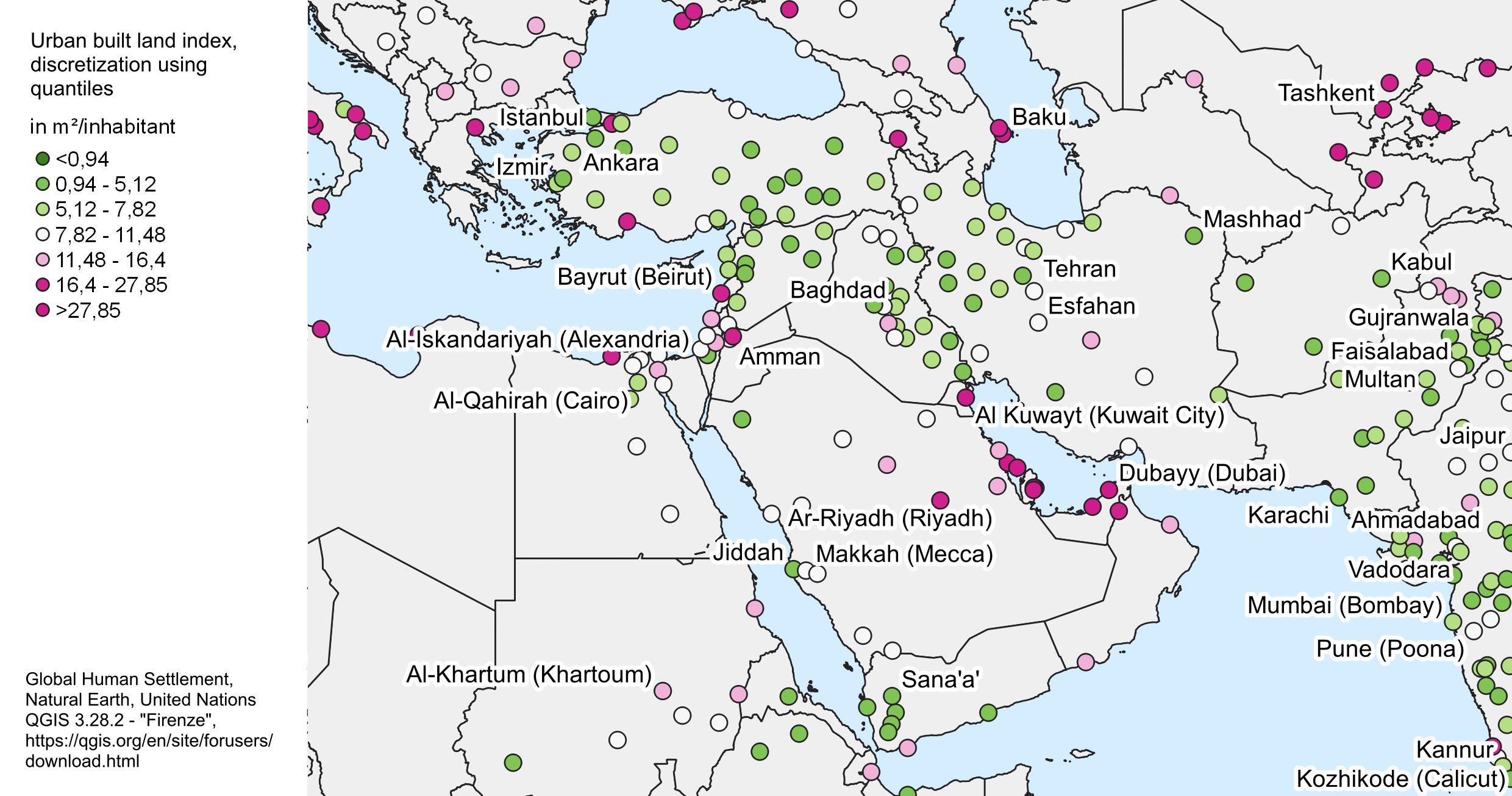}
\caption{Zooming on different parts of the global map of the Urban Built Land Index UBLI (Fig. 3 of the main text): South Asia, Europe, Middle East.}\label{zooms3}
\end{figure}

\subsection*{Relating national observations of the Urban Built Land Index to other variables}

In the main text, we map the UBLI at the global scale (Figure 3). Zooms on different continents are provided here on Figures \ref{zooms1}, \ref{zooms2} and \ref{zooms3}. We also describe in the main text the variations of the UBLI between countries, observed via the median UBLI at national scale. We relate the median UBLI (UBLI$_m$) to national economic and demographic variables on Figure 4. 

On Figure 4a), we relate it to the GDP pc of countries, obtaining a relationship UBLI$_m=5.74+5.28\times10^{-4}$GDPpc, which yields a R$^2$ of 0.906 (R=0.952) for the 24 countries having the highest numbers of cities. Figure \ref{corr} shows that the correlation between both variable is even higher if less countries are considered in this analysis. Indeed, we obtain R=0.963 when considering only the 14 countries having the highest numbers of cities, R=0.984 with only 10 countries, and R=0.99987 with only 3 countries. In our view, this shows that the relationship between UBLI and GDPpc at national scale is extremely strong, and that including more countries (with lower numbers of cities) in this analysis introduces more noise and weakens the correlation. Figure \ref{GDPmap} provides a global map of the GPDpc to be compared with the global map of the UBLI (Figure 3 of the main text). The same colors and discretization are used, and the national GDPpc is applied to all cities of the database. We know that this is not a correct geographical representation of this variable, but we provide it just for this comparison, which shows that both maps are very similar.

On Figure 4b), we relate UBLI$_m$ to mean household size (HHS), obtaining UBLI$_m=58\times$HHS$^{-1.4}$, with R$^2$=0.5 (R=-0.72).

 The GDP per capita is provided by the World Bank national accounts data, and OECD National Accounts data files, available at \href{https://data.worldbank.org/indicator/NY.GDP.PCAP.CD}{https://data.worldbank.org/indicator/NY.GDP.PCAP.CD} (last visited November 2024). The mean household size is provided by the United Nations Population  Division, within the Database on Household Size and Composition 2022, available at  \href{https://www.un.org/development/desa/pd/data/household-size-and-composition}{https://www.un.org/development/desa/pd/data/household-size-and-composition} (last visited November 2024). For the mean household size, the country data is collected and sorted according to the different sources which provide it. Then for each source, an autoregressive integrated moving average (ARIMA) is used to project the mean number of people per household in 2020 in each country. This methodology enables us to model and forecast this time series without being constrained by the disparities in data collection across different countries. Note  also that Saudi Arabia has no data for the household size indicator in our sources.

\end{document}